\documentclass[fleqn,usenatbib]{mnras}

\usepackage[T1]{fontenc}
\usepackage{graphicx}	
\usepackage{amsmath}	
\usepackage{amssymb}	
\usepackage[dvipsnames]{xcolor} 
\usepackage[normalem]{ulem} 
\usepackage{wasysym}
\usepackage{newtxtext,newtxmath} 

\defcitealias{Peroux&Howk20}{PH20}

\definecolor{darkgreen}{rgb}{0.0,0.75,0.0}
\definecolor{darkblue}{rgb}{0.0,0.0,0.5}
\definecolor{orange}{rgb}{0.75,0.5,0.0}
\definecolor{grey}{gray}{0.7}

\newcommand{\eg}[0]{$\textnormal{e.g. }$}
\newcommand{\ie}[0]{$\textnormal{i.e. }$}
\newcommand{\Msun}[0]{\,\textnormal{M}_{\textnormal{\astrosun}}}

\newcommand{\MetDen}[0]{$\rho\sub{met}$}

\newcommand{\MetDenNeu}[0]{$\rho\sub{met,neu}$}
\newcommand{\MetDenStars}[0]{$\rho_{\tn{met},*}$}
\newcommand{\MetDenTot}[0]{$\rho\sub{met,tot}$}
\newcommand{\SFRDen}[0]{$\rho\sub{\textsc{sfr}}$}
\newcommand{\HIDen}[0]{$\rho\sub{HI}$}

\newcommand{\MetNeu}[0]{$[\langle{}\tn{M/H}\rangle]\sub{neu}$}

\newcommand{\tn}[1]{\textnormal{#1}}
\newcommand{\sub}[1]{_{\textnormal{#1}}}

\newcommand{\error}[1]{\tn{\scriptsize{$\pm #1$}}}
\newcommand{\lgal}[0]{\textsc{L-Galaxies}}
\newcommand{\lgaltt}[0]{\textsc{L-Galaxies 2020}}

\newcommand{\logMm}[0]{\tn{log}_{10}(M_{*}/\Msun)}

\newcommand{\logten}[0]{\tn{log}_{10}}

\title[Cosmic neutral gas metal density]{Cosmic metal density evolution in neutral gas: insights from observations and cosmological simulations}

\author[Yates et al.]{
Robert M. Yates,$^{1}$\thanks{E-mail: r.m.yates@surrey.ac.uk}
C\'{e}line P\'{e}roux,$^{2,3}$
Dylan Nelson$^{4}$
\\\\
$^{1}$Department of Physics, University of Surrey, Stag Hill, Guildford, GU2 7XH, UK\\
$^{2}$European Southern Observatory, Karl-Schwarzschildstrasse 2, D-85748 Garching bei M{\"u}nchen, Germany\\
$^{3}$Aix Marseille Universit\'e, CNRS, LAM (Laboratoire d'Astrophysique de Marseille) UMR 7326, 13388, Marseille, France \\
$^{4}$Universit\"{a}t Heidelberg, Zentrum f\"{u}r Astronomie, Institut f\"{u}r theoretische Astrophysik, Albert-Ueberle-Str. 2, 69120 Heidelberg, Germany\\
}

\date{}
\pubyear{2021}

\begin{document}
\label{firstpage}
\pagerange{\pageref{firstpage}--\pageref{lastpage}}
\maketitle

\begin{abstract}
We contrast the latest observations of the cosmic metal density in neutral gas (\MetDenNeu{}) with three cosmological galaxy evolution simulations: \lgaltt{}, TNG100, and EAGLE. We find that the fraction of total metals that are in neutral gas is $<40$ per cent at $3\lesssim{} z \lesssim{} 5$ in these simulations, whereas observations of damped Lyman-$\alpha$ (DLA) systems suggest $\gtrsim{}85$ per cent. In all three simulations, hot, low-density gas is also a major contributor to the cosmic metal budget, even at high redshift. By considering the evolution in cosmic SFR density (\SFRDen{}), neutral gas density (\HIDen{}), and mean gas-phase metallicity (\MetNeu{}), we determine two possible ways in which the absolute \MetDenNeu{} observed in DLAs at high redshift can be matched by simulations: (a) the \SFRDen{} at $z\gtrsim{}3$ is greater than inferred from current FUV observations, or (b) current high-redshift DLA metallicity samples have a higher mean host mass than the overall galaxy population. If the first is correct, TNG100 would match the ensemble data best, however there would be an outstanding tension between the currently observed \SFRDen{} and \MetDenNeu{}. If the second is correct, \lgaltt{} would match the ensemble data best, but would require an increase in neutral gas mass inside subhaloes above $z\sim{}2.5$. If neither is correct, EAGLE would match the ensemble data best, although at the expense of over-estimating \MetNeu{}. Modulo details related to numerical resolution and \textsc{Hi} mass modelling in simulations, these incompatibilities highlight current tensions between key observed cosmic properties at high redshift.
\end{abstract}



\section{Introduction}\label{sec:Intro}
The evolution of the cosmic metal density (\MetDen{}) and cosmic star-formation rate density (\SFRDen{}) are key diagnostics for understanding galaxy evolution. Combined, they encode the relative significance of all the main physical processes driving galaxy evolution, including gas accretion, star formation, stellar evolution, metal enrichment, and outflows driven by supernovae (SNe) and active galactic nuclei (AGN).

Recently, observational studies of the metal content in and around galaxies have suggested that nearly all the metals in the Universe at $z\gtrsim{}2.5$ reside in regions dominated by neutral gas (\citealt{Peroux&Howk20}, hereafter PH20). These regions have \textsc{Hi} column densities of $N(\tn{\textsc{Hi}}) > 2\times{}10^{20}\,\tn{cm}^{-2}$ and are probed in absorption against bright background sources as damped Lyman-$\alpha$ (DLA) systems. DLAs are expected to trace the cold and warm neutral medium within galaxies \citep{wolfe2005,bird14,Rahmati+15,Peeples+19}, and do not suffer from the selection biases or metallicity calibration issues that hamper emission-based samples. An analysis of this DLA data by \citetalias{Peroux&Howk20} argues that the cosmic, dust-corrected, metal mass density in neutral gas (\MetDenNeu{}) is only a factor of a few lower than the total expected metal density in the Universe (\MetDenTot) at $z\sim{}2.5$, and constitutes $\gtrsim{}85$ per cent at $z\gtrsim{}4$. These estimates are higher than previously predicted from comparisons between DLAs and Lyman-break galaxies (LBGs) \citep{rafelski2014}.

Such a result is perhaps surprising, given that the presence of strong galactic winds in highly-star-forming galaxies and AGN hosts at high redshift (\eg{}\citealt{Adelberger+03,steidel10,Bischetti+19,Sugahara+19,Ginolfi+20,Spilker+20}) suggest an efficient removal of metals from galaxies. This material should enrich the circumgalactic medium (CGM) and intracluster medium (ICM) to high levels by $z\sim{}2$ (\eg{}\citealt{Mushotzky&Loewenstein97,Tozzi+03,Balestra+07,Anderson+09,Baldi+12,McDonald+16}), or at least be transferred to a hot, ionized phase, undetected by DLAs. The recent results on \MetDenNeu{} seem to be in tension with this scenario.

A precise census of the distribution of metals in the Universe at high redshift is difficult to obtain from current observations, as the full range of dominant contributors to the total metal budget have yet to be comprehensively measured (see \citetalias{Peroux&Howk20}). Therefore, in this work, we turn to large-volume cosmological galaxy evolution simulations to assess the apportionment of metals in and around galaxies back to high redshift from a theoretical perspective. This allows us to determine under which conditions \MetDenNeu{} dominates the universal metal budget above $z\sim{}2.5$, and what the consequences are for galaxy evolution in general. For this analysis, we  compare observational data with predictions from the \lgaltt{} semi-analytic model \citep{Henriques+20}, the EAGLE hydrodynamical simulation \citep{schaye15}, and the TNG100 magnetohydrodynamical simulation \citep{pillepich18b,springel18,nelson18a,naiman18,marinacci18}.

This paper is organised as follows: In Section \ref{sec:Obs}, we provide an overview of the observational data compiled and properties calculated by \citetalias{Peroux&Howk20} in order to form their analysis of metal densities. In Section \ref{sec:Sims}, we present the four cosmological-scale galaxy evolution models used in this work: \lgaltt{} `default model' (DM), \lgaltt{} `modified model' (MM), EAGLE, and TNG100. In Section \ref{sec:Results}, we discuss our findings, including the evolution of the cosmic metal, H\textsc{i}, and SFR densities, as well as neutral gas metallicities, in simulations and observations. In Section \ref{sec:future_prospects}, we outline the prospects for future observations at high redshift. Finally, in Section \ref{sec:Conclusions} we provide a summary of our conclusions.


\section{Observational data}\label{sec:Obs}
The so-called ``missing metals problem'' was originally posed some 20 years ago by \citet{pettini1999} and \citet{pagel1999} as an order of magnitude shortfall in the co-moving density of metals which have been measured compared with those expected to have been produced by the star formation activity seen in galaxies. At the time, when adding up all the metals which had been measured with some degree of confidence, findings showed that they accounted for no more than $\sim$10 per cent of what was expected to have been produced and released by $z=2.5$. Although, more recent studies suggest that this discrepancy is substantially reduced \citep[\eg{}][]{renzini1998, ferrara2005, bouche2005, bouche2006, bouche2007, shull2014}.

With the goal of reappraising the missing metals problem, \citetalias{Peroux&Howk20} reviewed this census by making an assessment of the cosmic metal mass density continuously with look-back time. Comprehensive estimates for the metal mass present in stars, the intragroup and intracluster medium (IGrM and ICM), highly and partially ionised gas (\ie{}Lyman Limit Systems, LLSs, and sub-DLAs), and neutral gas were made. Of particular interest for this study are the measurements made of the metal density in predominantly neutral gas as a function of redshift.

Absorption lines detected against bright background quasars offer the most compelling way to study the distribution, chemical properties, and metal mass budget of the dense gas in and around galaxies. In these quasar absorbers, the minimum gas density which can be detected is set by the brightness of the background source and the detection efficiency is independent of redshift. In emission, there is significant dispersion in the metallicities implied by different strong-line ratios and, more worryingly, between different calibrations of the same ratios \citep{Kewley&Ellison08}. Absorption techniques, on the other hand, directly count the number of atoms in a given phase of the gas. Several analyses report that abundances from emission and absorption vary by up to $\sim{}0.6$ dex (\eg{}\citealt{rahmani2016}, although see Section \ref{sec:ZDLA evo} below).

When probing regions dominated by neutral gas, the measurements trace only the dominant ionisation states of the metal element and hydrogen. Similarly, some of the metals are locked into dust grains so that observations of gas-phase metals might be inaccurate due to dust depletion. Historically, studies have attempted to avoid the difficulties associated with the dust bias by observing elements known to be only weakly affected by depletion, such as zinc (\eg{}\citealt{Vladilo+00}). More recently, a multi-element empirical method has been developed to correct homogeneously for dust depletion \citep{jenkins2009, decia2016, jenkins2017, decia2018a}. Such observations of the neutral phase of gas as traced by DLAs is available up to $z=5.3$, indicating only a mild evolution in mean metallicity over nearly a 13 Gyr timespan compared to other phases. After applying dust corrections, it appears observationally that the metal content in neutral gas dominates over more ionised phases at all redshifts. We also note that such multi-element, dust-corrected metallcities are, on average, 0.2-1.0 dex higher than uncorrected metallicity estimates using only Si or Fe lines. This emphasises both the uncertainty in single-ion DLA metallicity studies (especially at low redshift) and mitigates somewhat any differences in the way DLA metallicities are measured in simulations when comparing to data.

In parallel, an assessment of the total metal production by star formation is available from the compilation of \cite{Madau&Dickinson14}. By integrating \SFRDen{} over time, modulo the return fraction and assuming a yield of metal production, one can calculate the amount of metals expected to be produced by stars at any given epoch. A comparison of these results with the sum of the various contributors described above provides two independent estimates of the global metal budget. If the current direct measurements from DLAs of very large metal densities in neutral gas at high redshift are correct, then this would be sufficient to close the budget above $z\sim{}3$.

\section{Simulations}\label{sec:Sims}

In this section, we present the three cosmoloigcal galaxy evolution simulations used in this work. The key parameters and physical processes pertaining to metal evolution are also summarised in Table \ref{tab:sims}.

\begin{table*}
\centering
\begin{tabular}{lccccccccccc}
\hline \hline
Simulation & Box size & $m\sub{dm}$ & $m\sub{b}$ & $h$ & \multicolumn{3}{c}{Metal yields} & SFR & SN feedback \\
  & (Mpc/$h)^{3}$ & $10^{6}\Msun/h$ & $10^{6}\Msun/h$ & km/s/Mpc & AGB & SNe-Ia & SNe-II & formula & formula \\
\hline
\lgaltt{} & 96.1 & $7.7$ & -- & 0.673 & M01 & T03 & P98 & $\dot{\Sigma{}}_{*} = \alpha\Sigma\sub{H2}/t\sub{dyn}$ & $\dot{E}\sub{SN} = \eta{}\sub{SN}\epsilon\sub{SN}\dot{M}\sub{R}$ \\
EAGLE & 67.8 & $6.6$ & $1.2$ & 0.677 & M01 & T03 & P98 & $\dot{\Sigma{}}_{*} = A\Sigma\sub{gas}^{n}$ & $\dot{E}\sub{SN} = \chi{}\epsilon\sub{SN}\dot{M}_{*}$ \\
TNG100 & 75.0 & $5.1$ & $0.9$ & 0.677 & K10/D14/F14 & N97 & P98/K06 & $\dot{\rho}_{*} = \rho_{\rm c} / t_{*} $ & $\dot{E}\sub{SN} = \epsilon\sub{w}\dot{M}_{*}$  \\
\hline \hline
\end{tabular}
\caption{The key parameters and equations defining the three simulations considered in this work. More detail is provided in the text in Section \ref{sec:Sims}. The metal yield tables listed refer to the following works: \citeauthor{Marigo01} (2001, M01); \citeauthor{Thielemann+03} (2003, T03); \citeauthor{Portinari+98} (1998, P98); \citeauthor{karakas10} (2010, K10); \citeauthor{doherty14} (2014, D14); \citeauthor{fishlock14} (2014, F14); \citeauthor{nomoto97} (1997, N97); \citeauthor{kobayashi06} (2006, K06). For \lgaltt{}: $\alpha=0.06$ is the H$_2{}$-to-stars conversion efficiency, $\Sigma\sub{H2}$ is the H$_{2}$ surface mass density, $t\sub{dyn}=R\sub{cold}/V\sub{max}$ is the gas disc dynamical time, $\eta\sub{SN}=0.0149\Msun^{-1}$ is the number of SNe per mass, $\epsilon\sub{SN}=10^{51}$ erg is the energy released per SN, and $\dot{M}\sub{R}$ is the mass return rate by stellar winds and SNe (see \citealt{Henriques+20}). For EAGLE: $A=1.515\times{}10^{-4}\Msun\tn{/yr/kpc}^{2}$ is the gas-to-stars conversion efficiency, $n=1.4$ is the slope of the Kennicutt-Schmidt law \citep{Kennicutt98a}, $\Sigma\sub{gas}$ is the surface mass density for gas above a given metallicity-dependent density threshold \citep{schaye2010}, $\chi$ is the fraction of SN energy per mass used to drive galactic winds \citep{Springel&Hernquist03}, $\epsilon\sub{SN}=10^{51}$ erg is the energy released per SN, and $\dot{M}_{*}$ is the star formation rate (see \citealt{schaye15}). For TNG100: $\rho_{\rm c}$ is the cold gas density above a threshold value, $t_{*}$ is the (density-dependent) star formation timescale, which is taken to be proportional to the local dynamical time of the gas, and $\epsilon\sub{w}$ is the specific energy available for wind generation, which is tied to the SN-II energy per mass of stars formed \citep{pillepich18a}.}
\label{tab:sims}
\end{table*}

\subsection{\textsc{L-Galaxies 2020}}\label{sec:LGals}

\lgaltt{}\footnote{\url{https://lgalaxiespublicrelease.github.io}} \citep{Henriques+20} is a semi-analytic model of galaxy evolution, built to run on the dark matter (DM) subhalo merger trees of N-body DM simulations. It improves on previous versions of the \lgal{} model by including prescriptions for H$_{2}$ formation \citep{Fu+10}, detailed chemical enrichment from supernovae (SNe) and stellar winds \citep{Yates+13}, and radially-resolved gas and stellar galactic discs \citep{Fu+13}. In this work, \lgaltt{} is run on the \textsc{Millennium-II} simulation \citep{Boylan-Kolchin+09} with box side length $96.1\,h^{-1} \sim{} 143\,\tn{cMpc}$ and DM particle mass $7.7\times{}10^{6}\,h^{-1}\Msun$, in order to be comparable to the volumes of the hydrodynamical simulations which we also study (see below).

The key free parameters in \lgaltt{}, such as the efficiency of conversion of H$_{2}$ into stars, the efficiency of supermassive black hole (SMBH) feeding, and others, are simultaneously calibrated following an MCMC formalism \citep{Henriques+09,Henriques+15} in order to match the observed stellar mass functions and red fractions at $z=0-2$ and the observed \textsc{Hi} mass function at $z=0$. Further information on the baryonic physics implemented into \lgaltt{} can be found in the supplementary material.\footnote{\url{https://lgalaxiespublicrelease.github.io/Hen20_doc.pdf}} Following the method of \citet{Angulo&White10} and \citet{Angulo&Hilbert15}, the \lgaltt{} cosmology was rescaled to that of the \textit{Planck 2013} survey, as described by \citet{Planck14}: $\Omega_{\Lambda,0}=0.685$, $\Omega\sub{m,0}=0.315$, $\Omega\sub{b,0}=0.0487$, $\sigma_8=0.826$, $n\sub{s}=0.96$, $h=0.673$.

One of the main advantages of semi-analytic models is that the prescriptions governing baryonic physics can be easily adapted. Therefore, in this work, we focus on two versions of the \lgaltt{} model -- the `default model (DM)' and the `modified model (MM)'. The DM was first presented in \citet{Henriques+20} and assumes 70 percent of the metal released by SNe is immediately mixed with the local ISM, before some is driven out of galaxies through SN-driven galactic outflows. This equates to direct CGM enrichment efficiencies for SNe and AGB winds of $f\sub{SNII,hot} = 0.3$, $f\sub{SNIa,hot} = 0.3$, and $f\sub{AGB,hot} = 0.0$. The MM was first presented in \citet{Yates+21a} and assumes up to 90 per cent of the metal released by SNe is allowed to directly enrich the CGM surrounding galaxies, without first mixing with the ISM, \ie{}$f\sub{SNII,hot} = 0.9$, $f\sub{SNIa,hot} = 0.8$, and $f\sub{AGB,hot} = 0.25$. These two versions of \lgaltt{} therefore nicely bracket the likely range of direct CGM enrichment efficiencies present in real galaxies.

The stellar yield tables used in both versions of \lgaltt{} are given in Table \ref{tab:sims}. These are the same as those used in the EAGLE simulation.

\subsection{EAGLE}\label{sec:EAGLE}

The EAGLE simulation project\footnote{\url{eagle.strw.leidenuniv.nl}} \citep{crain15, schaye15} is a series of cosmological, hydrodynamical simulations, all run with a modified version of the \textsc{Gadget-3} code \citep{spr05}. Here we exclusively use the largest volume `Ref-L100N1504' simulation, which has a $67.8\,h^{-1} \sim{}100\,\tn{cMpc}$ side-length box containing $2 \times 1504^3$ dark matter plus baryonic particles. As a result, the (initial) particle masses are $1.2\times10^6\,h^{-1}\Msun$ and $6.6\times10^6\,h^{-1}\Msun$, for gas and DM, respectively, while the collisionless gravitational softening length is fixed at 0.7\,physical kpc at late times ($z<2.8$). The EAGLE cosmology is based on that from the \textit{Planck 2013} survey, as described by \citet{Planck14_I}: $\Omega_{\Lambda,0}=0.693$, $\Omega\sub{m,0}=0.307$, $\Omega\sub{b,0}=0.04825$, $\sigma_8=0.829$, $n\sub{s}=0.9611$, $h=0.6777$.

The EAGLE model for galaxy formation physics includes a comprehensive description of the most important processes governing galaxy growth and evolution. This incorporates star formation, whereby gas above a (metallicity-dependent) threshold density is stochastically converted into stars; stellar evolution tracks the abundances of eleven individual elements, following massive and intermediate-mass stars; radiative cooling and heating are based on these individual elemental abundances, assuming photoionization equilibrium with an optically thin ionizing background radiation field \citep{wiersma09}; supermassive black holes are seeded in all sufficiently massive halos, and subsequently grow by accreting gas from their surroundings; stellar feedback as well as supermassive black hole feedback are both treated via a bursty thermal energy release \citep{dallavecchia12}.

\subsection{TNG100}\label{sec:TNG100}

The IllustrisTNG project\footnote{\url{www.tng-project.org}} \citep{naiman18, pillepich18b, nelson18a, marinacci18, springel18} is a series of large-volume cosmological gravo-magnetohydrodynamics (MHD) simulations, run with a comprehensive model for the physics driving galaxy formation \citep{weinberger17,pillepich18a}. The TNG project encompasses three uniform volume boxes: TNG50, TNG100, and TNG300, all run with the exact same physical model, albeit at different numerical resolutions. Here we exclusively use the intermediate TNG100 run, as a compromise between the high-resolution TNG50 realization \citep{pillepich19,nelson19b} and the large-volume TNG300 run. In particular, TNG100 includes 1820$^3$ dark matter particles, plus a roughly equal number of baryonic mass elements spanning gas, stars, and supermassive black holes, within a $75\,h^{-1} \sim{}110\,\tn{cMpc}$ side-length box.

TNG100 has a baryon mass resolution of $9.4\times{}10^5\,h^{-1}\Msun$, a DM mass resolution of $5.1\times{}10^6\,h^{-1}\Msun$, a collisionless softening of 0.74 physical kpc at $z=0$, and a minimum gas softening of 184 comoving parsecs. Its cosmology is taken from \textit{Planck 2015} data \citep{planck2015_xiii}, adopting $\Omega_{\Lambda,0}=0.6911$, $\Omega\sub{m,0}=0.3089$, $\Omega\sub{b,0}=0.0486$, $\sigma_8=0.8159$, $n\sub{s}=0.9667$ and $h=0.6774$. Halos and subhalos (galaxies) are identified with the \textsc{Subfind} algorithm \citep{spr01}, as in EAGLE.

TNG uses the \textsc{Arepo} code \citep{spr10} and includes a model for the key processes which regulate the formation of galaxies. Most importantly, feedback from both supernovae and supermassive black holes \citep[see][for details]{weinberger17,pillepich18a}. The production of metals and the assumed stellar nucleosynthetic yields are taken from a combination of multiple stellar evolution codes, as shown in Table \ref{tab:sims}, noting that these yield tables differ from those in the original Illustris simulation \citep[see][]{pillepich18a}.

\subsection{Differences between the simulations}\label{sec:sim differences}

In addition to differing choices in the stellar yield sets used, the three simulations considered here also differ in the number of chemical elements tracked. Both \lgaltt{} and EAGLE track 11 chemical elements: H, He, C, N, O, Ne, Mg, Si, S, Ca, Fe. The last nine of these (\ie{}excluding H and He) constitute $\sim{}98$ per cent of the total metal mass in the photosphere of the Sun \citep{Asplund+09}, indicating that their sum provides a good approximation to the total metal mass. TNG100 only explicitly tracks 9 chemical elements: H, He, C, N, O, Ne, Mg, Si, Fe, the last seven of which constitute $\sim{}95$ per cent of the total metal mass in the Sun's photosphere. However, this does not make a significant difference to abundances in TNG100, as the SN-II yields (which dominate the production of the most abundant metals) are also renormalised to match the total mass ejected from the \citet{Portinari+98} tables used in \lgaltt{} and EAGLE (see \citealt{pillepich18a}).

The SN feedback models in \lgaltt{}, EAGLE, and TNG100 also differ in a number of substantial ways. This impacts the metal enrichment of gas in and around galaxies. The efficiency with which SN feedback drives galactic outflows can be quantified by the mass loading factor, $\eta = \dot{M}\sub{w}/\dot{M}_{*}$, where $\dot{M}\sub{w}$ is the mass outflow rate and $\dot{M}_{*}$ is the star formation rate. This factor essentially represents the efficiency of gas expulsion for a given amount of available SN energy (see Table \ref{tab:sims}).

In EAGLE, $\eta$ decreases with cosmic time at fixed $\dot{M}_{*}$ \citep{Mitchell+20a}, due to the fraction of SN energy made available for outflows being inversely proportional to metallicity \citep{Furlong+15}. This means that metals can be more efficiently accumulated in the ISM at later times, playing an important role in the rate of metal enrichment, as shown in Section \ref{sec:ZDLA evo}.

In TNG100, $\eta$ is typically higher than in EAGLE \citep{nelson19b}, indicating lower metal retention. However, an increased \SFRDen{} can outweigh this effect, through the greater production of metals overall, as discussed in Section \ref{sec:SFRD evo}. Also, the redshift evolution of $\eta$ in TNG100 is somewhat smoother over time \citet{nelson19b}, at least for galaxies with $\logMm{}<10.5$ where AGN feedback does not play a significant role \citep{ayromlou20}.

In \lgaltt{}, $\eta{}$ values are similar to those in EAGLE at $z=0$ \citep{Yates+21a}. However, like TNG100, they are not directly dependent on metallicity, allowing for a more efficient retention of metals in the ISM at high redshift than in EAGLE. For the DM version of \lgaltt{}, this is enough to return high ISM metallicities in low-mass galaxies at early times. For \lgaltt{} MM however, this effect is strongly offset by the increased metal ejection efficiency (see Section \ref{sec:LGals}), which greatly reduces ISM metallicities at high redshift, despite this version of the simulation having lower overall $\eta{}$ \citep{Yates+21a}.

The star formation models in the three simulations are more similar to each other (see Table~\ref{tab:sims}). \lgaltt{} adopts an H$_{2}$-dependent SFR law, made  possible because the partitioning of ISM gas into \textsc{Hi} and H$_{2}$ is modelled on the fly in the simulation \citep{Fu+10}. EAGLE and TNG100 instead follow the prescription developed by \citet{Springel&Hernquist03}, which relates the star formation rate density to the total density of gas above a certain density threshold of $n_{\rm H} \simeq 0.1 \rm{cm}^{-3}$ \citep{vog13,schaye15}. Both of these approaches reproduce the observed Kennicutt-Schmidt relation \citep{Kennicutt98a} for star-forming galaxies.

Finally, the small differences in cosmological parameters assumed between the simulations should only have a negligible impact on their relative results \citep{Angulo&White10,Guo+13,Henriques+15}, especially when compared to differences in the implementation of baryonic physics.


\section{Results}\label{sec:Results}

\subsection{Metal density evolution}\label{sec:Met den evo}

In this section, we present the main focus of this work -- a comparison between the cosmic metal density evolution from observations and the \lgaltt{}, EAGLE, and TNG100 simulations.

\subsubsection{Measuring densities}\label{sec:Measuring densities}

\begin{figure*}
\centering
 \includegraphics[angle=0,width=0.99\linewidth]{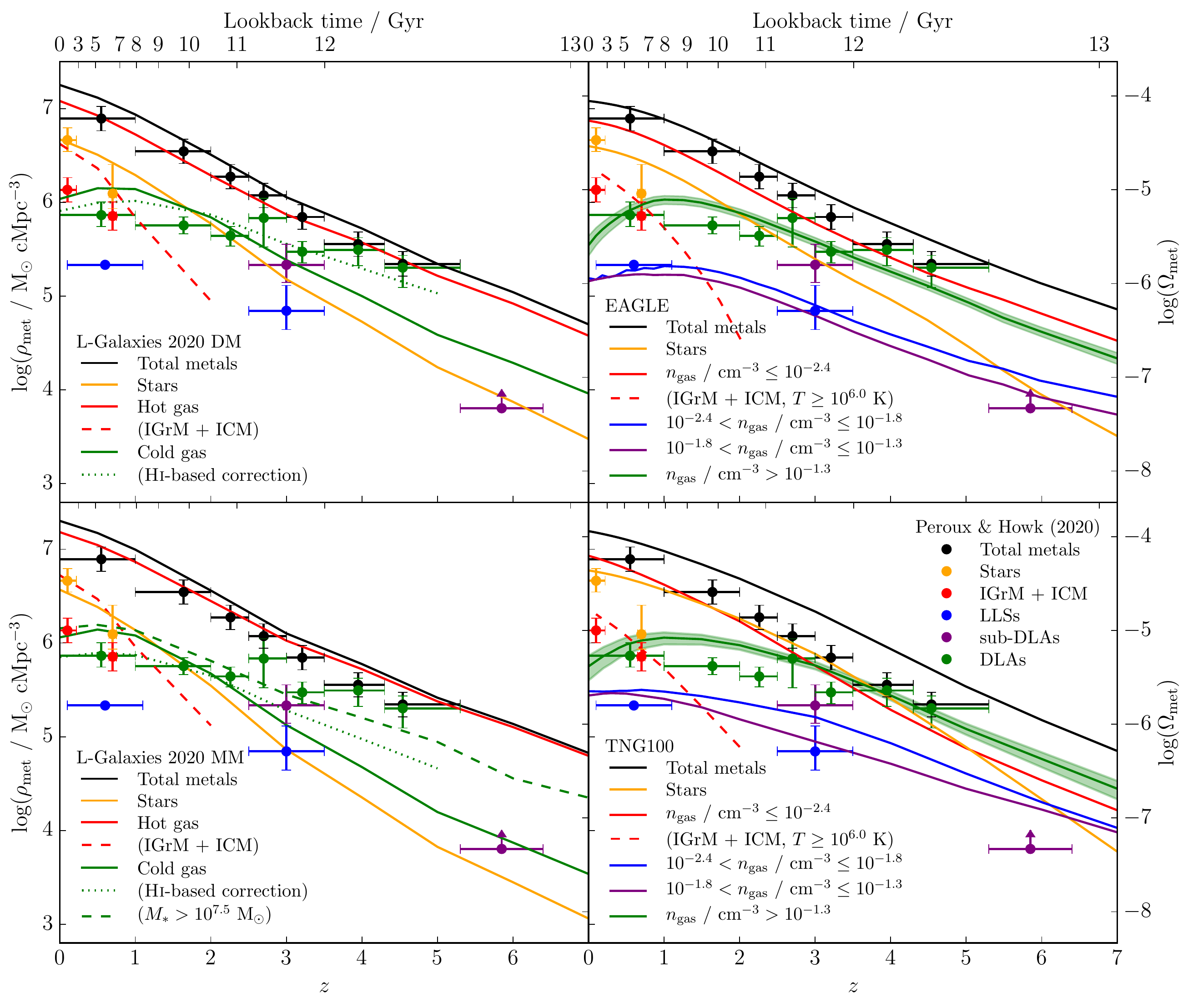} 
 \caption{Evolution of the cosmic metal density in various phases in and around galaxies from $z=7$ to 0. Each panel represents one of the four galaxy evolution models considered here (see legends). Lines represent model results, with green shaded regions representing the uncertainty in the DLA density threshold for EAGLE and TNG100 (see Section \ref{sec:Measuring densities}). Points (shown in all four panels) represent observational results from \citetalias{Peroux&Howk20}. Dotted green lines represent \MetDenNeu{} for \lgaltt{} when corrected with the \textsc{Hi} density model described in Section \ref{sec:Bary den evo}. Dashed green line represents \MetDenNeu{} when only considering galaxies with $\logMm{} > 7.5$ in \lgaltt{} MM (see Section \ref{sec:ZDLA evo}).}
 \label{fig:Metal_density_evo}
\end{figure*}

For the simulations, metal densities are calculated at any given redshift as the total metal mass in that particular phase divided by the total comoving volume of the simulation box,
\begin{equation}\label{eqn:metalDensity_theo}
\rho\sub{met,theo} = \frac{\sum{M\sub{metals}}}{V\sub{tot}}\ .
\end{equation}

Observationally, neither $\sum{M\sub{metals}}$ nor $V\sub{tot}$ are readily measurable, so subtly different approaches are taken to infer metal densities. When estimating the total metal density in neutral gas, the following is used
\begin{equation}\label{eqn:metalDensity_obs}
\rho\sub{met,neu,obs} = \rho\sub{HI+He}\cdot{}\langle{}M/H\rangle{}\sub{\textsc{dla}}\ \ \ ,
\end{equation}

where $\rho\sub{HI+He}$ is the universal baryon density of \textsc{Hi} plus helium, as probed by statistical \textsc{Hi} surveys at various redshifts (see \citetalias{Peroux&Howk20} and Section \ref{sec:Bary den evo}), and $\langle{}M/H\rangle{}\sub{\textsc{dla}}$ is the \textsc{Hi}-weighted mean metal-to-hydrogen mass ratio from DLAs (see \citetalias{Peroux&Howk20}, section 3.4). Eqns. \ref{eqn:metalDensity_theo} and \ref{eqn:metalDensity_obs} return essentially the same result (modulo small differences between the mass of hydrogen, \textsc{Hi}+He, and total neutral gas) in the limit that DLAs are unbiased tracers of neutral gas in the Universe. However, if this is not the case, these two calculations can differ, as discussed in Section \ref{sec:ZDLA evo}. Previous works have argued that, when selecting DLAs from simulations along random sightlines, and modeling \textsc{Hi} disc sizes and densities, the cosmic neutral gas density inferred is similar to that obtained by considering all \textsc{Hi} in the simulation box \citep{Berry+16,Di_Gioia+20}.

We highlight here that the cosmic mass densities given by equations Eqns. \ref{eqn:metalDensity_theo} and \ref{eqn:metalDensity_obs} are essentially `normalised' total metal masses. This is because the volume considered is always the total comoving volume of the Universe (explicitly in the case of Eqn. \ref{eqn:metalDensity_theo}, and implicitly in the case of Eqn. \ref{eqn:metalDensity_obs}). The actual volume filling fraction of any particular phase does not enter. The equivalent, dimensionless quantity, $\Omega = \rho/\rho\sub{crit,0}$, is therefore sometimes preferred in the literature, where $\rho\sub{crit,0}=3\tn{H}_{0}^{2}/8\pi{}\tn{G}=1.36\times10^{11}\Msun\,\tn{cMpc}^{-3}$ is the critical mass density of the Universe at $z=0$. This quantity is shown as the second y-axis in all relevant figures in this work.

The various baryonic phases are distinguished in different ways in the simulations. For \lgaltt{}, baryons within haloes are split into seven distinct components: SMBH, bulge stars, disc stars, halo stars, cold gas (\ie{}ISM), hot gas (\ie{}CGM), and an ejecta reservoir. The cold gas component within galaxies (which is tuned to reproduce the observed \textsc{Hi} mass function at $z=0$, see \citealt{Henriques+20}) is assumed to represent neutral gas in this work. For EAGLE and TNG100, gas particles/cells are assigned to phases according to their volumetric density. Density ranges have been chosen to roughly replicate the column density designations given to DLAs, sub-DLAs, LLSs, and sub-LLSs in observations (see Fig. \ref{fig:Metal_density_evo}). Therefore, dense neutral gas is assumed to be represented by material with $n\sub{gas} > 10^{-1.3}\,\tn{cm}^{-3}$. This corresponds to the threshold column density for DLAs of $N(\textsc{Hi}) = 10^{20.3}\,\tn{cm}^{-2}$ at $z=3$ in the EAGLE precursor simulations studied by \citet{rahmati13} (see also \citetalias{Peroux&Howk20}, fig. 1c). Because the relation between volumetric and column densities is not exact, a factor of 2 uncertainty (\ie{}a range of thresholds from $10^{-1.6}$ to $10^{-1.0}\,\tn{cm}^{-3}$) is accounted for in this work, matching the $1\sigma$ spread in $n\sub{gas}$ found by \citet{rahmati13} at $N(\textsc{Hi}) = 10^{20.3}\,\tn{cm}^{-2}$. This uncertainty in EAGLE and TNG100 results is indicated by shaded regions in all relevant figures.

\subsubsection{Apportionment of metals}\label{sec:Comparing obs and sims}

Fig. \ref{fig:Metal_density_evo} shows the cosmic metal density evolution for various phases in and around galaxies. Each panel shows results from a different simulation considered here (lines), with \citetalias{Peroux&Howk20} observational data shown in all panels (points). 

At low redshift ($z\lesssim 2.5$), all four simulations reproduce the observed partition of metals among various phases reasonably well. For example, EAGLE and \lgaltt{} (both the DM and MM) match the steep increase in \MetDenStars{} observed from $z\sim0.75$ to the present day (orange). EAGLE and TNG100 also roughly reproduce the normalisation of the metal density in neutral gas at $z\sim{}0.5$ from DLAs (green), as well as the total gas metal density at $z\sim{}1$ inferred from SED modelling of present-day galaxies \citep{Bellstedt+20}.

Importantly, the metals found in gas with $n\sub{gas} \leq{} 10^{-2.4}\,\tn{cm}^{-3}$ in EAGLE and TNG100, or equivalently in the hot CGM in \lgaltt{}, make up a significant fraction of the total metal budget (solid red lines). The dominance of this phase in the simulations highlights its importance when assessing the ``missing metals problem''. Such material is roughly equivalent to sub-LLS gas with $N(\tn{\textsc{Hi}}) < 1.6\times{}10^{17}\,\tn{cm}^{-2}$ in observations, as traced by the Ly$\alpha$ forest. The full complement of this low-density phase is difficult to measure observationally, especially at high redshift, so is not included in the observational data set shown in Fig.~\ref{fig:Metal_density_evo}. However, the sub-component of this phase from the IGrM+ICM at low-redshift is shown (red points). This can be compared to the dashed red lines from the simulations, which represents hot gas in within subhaloes of mass $\tn{log}(M_{200}/\Msun) \geq{} 13.0$. In all three simulations, this sub-component has a relatively small contribution to the total low-density gas in the universe, which is predominantly found in the subhaloes of field galaxies. In TNG100, the dominant metal fraction is shared roughly equally between this low-density gas and stars, highlighting the large amount of metals locked-up in stars in that simulation.

At high redshift ($z\gtrsim{}2.5$), the differences between the simulations and observations are more pronounced. For example, EAGLE appears to under-estimate \MetDenNeu{} by $0.27$ dex at $z\sim{}4$ compared to observations, whereas the \lgaltt{} DM and MM under-estimate this much more significantly, by $0.49$ dex and $0.82$ dex, respectively. In turn, TNG100 is able to reasonably match the observed \MetDenNeu{} from DLAs at high redshift. However, in order to do this, the total metal budget (black line), is higher than that inferred from either observations or the other simulations considered here. The cause of this discrepancy is discussed in Section \ref{sec:SFRD evo}.

Overall, the observational claim that nearly all metals at high redshift are in neutral gas is not reproduced in any of the simulations. While DLA observations suggest that $\sim{}87$ per cent of the total expected metal budget at $z\sim{}4$ is in neutral gas, less than 40 per cent is in neutral gas in the simulations considered here at the same redshift (19 per cent for \lgaltt{} DM, 8 per cent for \lgaltt{} MM, 30 per cent for EAGLE, and 32 per cent for TNG100). Hot gas, and stars in the case of TNG100, also have a major contribution in the simulations. This material remains largely unobserved at these redshifts.

\subsubsection{Biases in observations and simulations}\label{sec:Biases in observations and simulations}

When considering the accuracy of \MetDenNeu{} in observations, we note that the dust corrections applied to $\langle{}M/H\rangle{}\sub{\textsc{dla}}$ measurements are significant (up to $\sim{}0.5$ dex for iron at high redshift), and can, in principle, lead to over-estimates. However, the dust correction technique utilised by \citet{De_Cia+18} for this data set is relatively well constrained, relying on multiple metal absorption lines. This means that the dust corrections applied here are likely to be reliable. A lack of highly dust-obscured systems in the sample can also pose an issue, however this would lead to an \textit{under}-estimate of \MetDenNeu{}, rather than an over-estimate, therefore \textit{increasing} the discrepancy between the observations and simulations. Dust is therefore unlikely to fully explain the discrepancy seen in Fig. \ref{fig:Metal_density_evo}.

When considering the accuracy of metal density measurements in simulations, we note that mass resolution limits could lead to inaccurate estimates for very low mass systems (see \eg{}\citealt{Di_Gioia+20}). However, the amount of metals residing within sub-resolution systems would have to be considerably under-estimated for this to fully account for the discrepancies seen in Fig. \ref{fig:Metal_density_evo}. Nonetheless, such mass resolution issues do significantly affect model \textsc{Hi} mass densities, as discussed in Section \ref{sec:Bary den evo}.

Finally, in contrast to the argument that model galaxies \textit{below} a certain mass should contribute more metals, an alternative argument is that the observed DLA samples are only probing systems \textit{above} a certain mass. We find that when limiting the minimum stellar mass assumed for DLA hosts in the simulations (and therefore using Eqn. \ref{eqn:metalDensity_obs} rather than Eqn. \ref{eqn:metalDensity_theo} to calculate \MetDenNeu{}), the correspondence with the observed \MetDenNeu{} from DLAs is much improved for \lgaltt{} MM  (green dashed line in Fig. \ref{fig:Metal_density_evo}). We discuss this possibility and other complications in the estimation of \MetDenNeu{} from DLAs in Section \ref{sec:ZDLA evo}.

In conclusion, we find that none of the cosmological simulations considered here agree with the observational finding that the majority of metals in the Universe at high redshift are in the neutral gas phase. Instead, all four simulations suggest that low-density hot gas also has a significant contribution. However, the absolute amount of metals observed in DLAs can be matched by simulations, but only under certain conditions outlined in the following sections.

\subsection{Star formation rate density evolution}\label{sec:SFRD evo}

When assessing the metal densities seen at high redshift, it is important to also consider the evolution of the cosmic SFR density (\SFRDen{}). This is because, to first order, \SFRDen{} controls the rate of metal production in the Universe and the distribution of these metals throughout galaxies and their environments via outflows.

\begin{figure}
\centering
 \includegraphics[angle=0,width=0.99\linewidth]{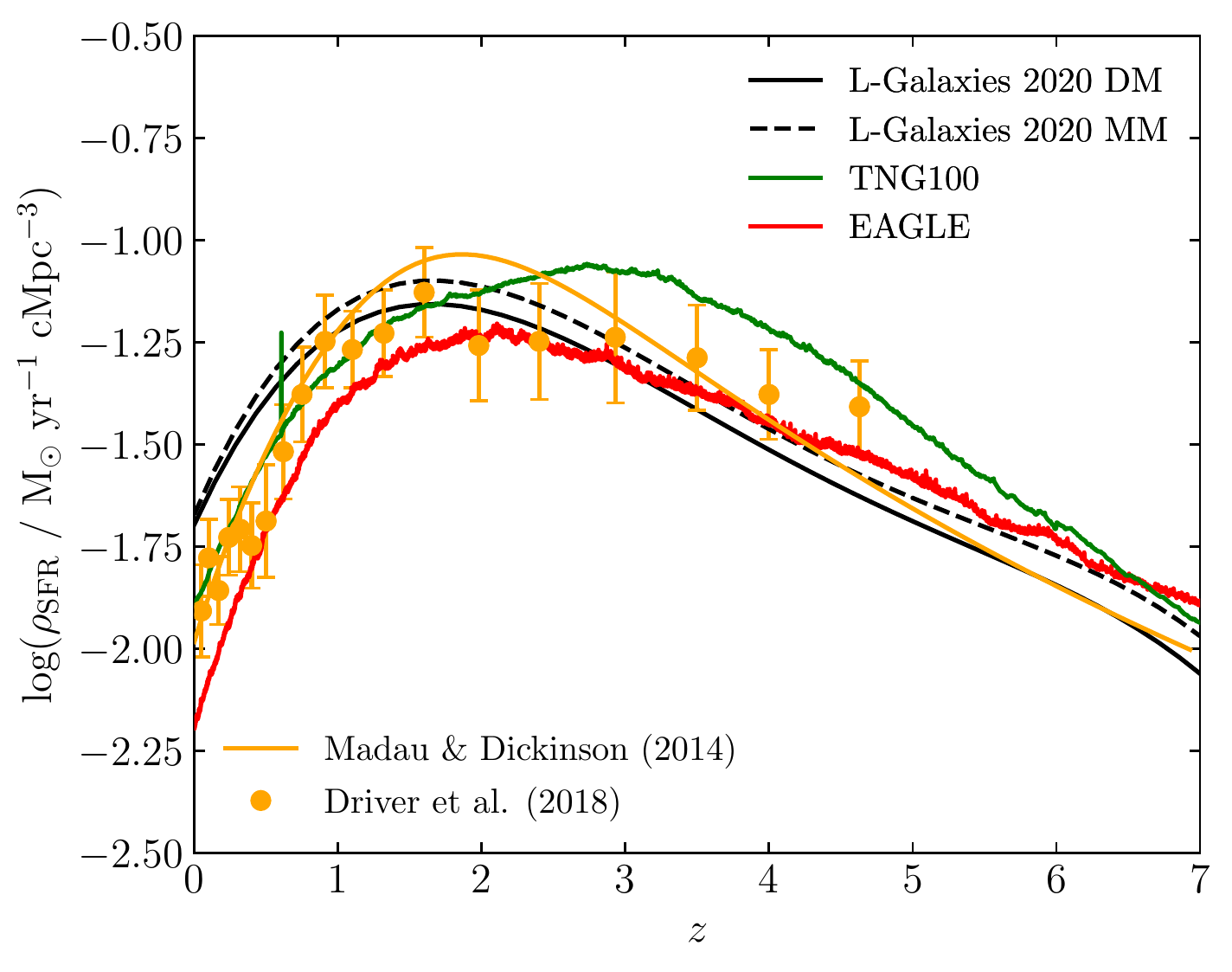}
 \caption{Cosmic star formation rate density versus redshift for the four galaxy evolution models considered here. Observational data from \citeauthor{Madau&Dickinson14} (2014, orange line) and \citeauthor{Driver+18} (2018, orange points) are also shown for comparison. We contrast four simulations: \lgaltt{} SM (solid black), \lgaltt{} MM (dashed black), TNG100 (green), and EAGLE (red).}
 \label{fig:SFRD_evo}
\end{figure}

Fig. \ref{fig:SFRD_evo} shows the evolution of \SFRDen{} for the four simulations considered here, along with two observational compilations from \citet{Madau&Dickinson14} (orange line) and \citet{Driver+18} (orange points). Both these studies incorporate data spanning rest-frame wavelengths from the far infrared (FIR, at lower redshifts) to the far ultraviolet (FUV). Comparing the observed \SFRDen{} with theoretical models in this way is difficult, as many systematic uncertainties exist, including the importance of obscured star formation, biased galaxy samples, resolution effects, and cosmic variance. Nonetheless, we present here a face-value comparison, and discuss the complexities involved.

The \lgaltt{} DM and MM match the observed \SFRDen{} shown in Fig. \ref{fig:SFRD_evo} quite well at high redshift, but over-estimate the total amount of star formation below $z\sim{}1$ (when run on the \textsc{Millennium-II}). Conversely, EAGLE slightly over-estimates the observed \SFRDen{} at $z\gtrsim{}5$, but under-estimates it by over $0.2$ dex at $z\lesssim{}2$ (see also \citealt{Furlong+15,Katsianis+17b,Driver+18}). TNG100 has comparable (or perhaps too \textit{low}, \citealt{donnari19}) star formation rates at $z\sim2$, and is also consistent with 3D-HST data at $z \sim 1.5$ \citep{Nelson+21}. However, TNG100 also shows the largest difference at high redshift, with significantly more star formation at $3 \lesssim z \lesssim{} 5$ than inferred by \citet{Madau&Dickinson14} and \citet{Driver+18}, and an earlier peak at $z\sim{}3$ (see also \citealt{Bellstedt+20,Zhao+20}). Cosmic stellar mass densities, $\rho_{*}$, are also higher than observed. This enhanced star formation relative to the other simulations at $z\sim{}3-5$ is the cause of the higher metal densities seen in TNG100 in Fig. \ref{fig:Metal_density_evo}.

\citet{Furlong+15} discuss how a combination of burstier star formation and reduced AGN feedback efficiency would enable higher SFRs and therefore increased \SFRDen{} in EAGLE, particularly around the peak of cosmic star formation at $z\sim{}2$. However, we note that this could also lead to an over-estimate of \MetDenNeu{} compared to low-redshift DLA observations, which are already slightly exceeded by EAGLE (see Fig. \ref{fig:Metal_density_evo}). Similarly, a reduction in the star formation efficiency at high redshift in TNG100 (particularly in the progenitors of the most-massive galaxies by $z=0$, see \citealt{Bellstedt+20}) would lead to a better agreement with current \SFRDen{} inferences from FUV data, but would also cause a greater under-estimate of \MetDenNeu{} at high redshift.

Therefore, a comparison between Fig. \ref{fig:Metal_density_evo} and Fig. \ref{fig:SFRD_evo} suggests an incompatibility between the currently observed \MetDenNeu{} and \SFRDen{} from FIR-FUV data. This incompatibility is reminiscent of that noted between some \SFRDen{} and $\rho_{*}$ estimates, which is now largely understood (\eg{}\citealt{Wilkins+19,leja20,Koushan+21}).

When assessing this incompatibility, we note that current observational determinations of \SFRDen{} using FUV data could under-estimate the amount of star formation in the high-redshift Universe (\eg{}\citealt{Bellstedt+20,Asada+21}). Significant uncertainties in \SFRDen{} are caused by limited sky coverage and the incompleteness of samples, which make large, uncertain extrapolations of the galaxy luminosity function necessary (\eg{}\citealt{Madau&Dickinson14,Katsianis+17b}). Dusty star-forming galaxies can also be missed when selecting and/or analysing galaxies at shorter wavelengths (\eg{}\citealt{Rowan-Robinson+16,Wang+19}). Recent studies of ALMA data for a small number of systems from the MORA and ALPINE surveys suggest that such obscured star formation could contribute between 17 and 35 per cent of the total \SFRDen{} at $z\sim{}5$ (\eg{}\citealt{Gruppioni+20,Zavala+21}). Inferences from GRBs also suggest a higher \SFRDen{} at $z>5$ \citep{Kistler+09}, and a recent study by \citet{Matthews+21} utilising radio data suggests a systematically higher \SFRDen{} at $z\lesssim{}3.5$. Importantly, at $z\sim{}4$, where TNG100 shows the biggest discrepancy with observations in Fig. \ref{fig:SFRD_evo}, \citet{Katsianis+20} find that rest-frame UV data could under-predict SFRs in more massive galaxies by $\sim{}0.13$ dex. Although, we note that other studies have shown that FIR-FUV measurements can instead \textit{over-estimate} the true \SFRDen{} (\eg{}\citealt{Utomo+14,Katsianis+17b,leja19,Katsianis+20}).

A systematic bias in the observed \SFRDen{} will also affect the total expected metal density, \MetDenTot{} (black points in Fig. \ref{fig:Metal_density_evo}). \citetalias{Peroux&Howk20} determine this quantity by integrating the \SFRDen{} from \citet{Madau&Dickinson14} over time, assuming a fixed mass return fraction and metal yield. It is therefore sensitive not only to changes in stellar lifetimes, remnant masses, and metal yields with redshift, but also the accuracy of SFR measurements. Therefore, an increase in \SFRDen{} at $3\lesssim{}z\lesssim{}5$ would cause a corresponding increase in \MetDenTot{}, reducing the fraction of total metals found in neutral gas. This would reduce the discrepancy between observations and simulations discussed in Section \ref{sec:Met den evo}.

However, in addition to observational uncertainties, biases in simulations due to their limited resolution and volumes should also be considered (\citealt{schaye2010,pillepich18b}). We find that the \SFRDen{} obtained from the larger TNG300 simulation is $\sim{}0.2$ dex \textit{lower} than seen in TNG100, bringing it into closer agreement with FIR-FUV observations at $z\sim{}3$, although worse agreement below $z\sim{}2$ (see also \citealt{Zhao+20}). Similarly, \SFRDen{} in \lgaltt{} when run on the larger \textsc{Millennium-I} simulation is lower than when run on \textsc{Millennium-II} (see also \citealt{Henriques+20}). These differences are predominantly due to lower resolution in the larger simulations, rather than cosmic variance issues in the smaller simulations \citep{genel14,Zhao+20}. Lower resolution leads to an under-abundance of low-mass galaxies, which in turn leads to an under-estimation of \HIDen{} and \SFRDen{}. For example, \citet{Spinelli+20} find that haloes with $M_{200} \lesssim{} 10^{11}\Msun$ contribute the majority of \HIDen{} above $z\sim{}2.5$ in GAEA (see also \citealt{lagos2014}), and \citet{villaescusa-navarro2018} find an under-estimate in \HIDen{} at high redshift in TNG300 relative to TNG100. Likewise, \SFRDen{} in both \textsc{Illustris} and EAGLE is dominated by lower-mass galaxies at high redshift \citep{genel14,Katsianis+17b}.

We therefore conclude that one avenue for cosmological simulations to reproduce the observed \MetDenNeu{} at high redshift is to produce stars (and therefore metals) at a higher rate than is expected from current FIR-FUV observations of \SFRDen{}. This implies a tension between the currently-observed \MetDenNeu{} and \SFRDen{}, and highlights the importance of considering both properties when studying galaxy evolution.

\subsection{Neutral gas density evolution}\label{sec:Bary den evo}

\begin{figure*}
\centering
 \includegraphics[angle=0,width=0.99\linewidth]{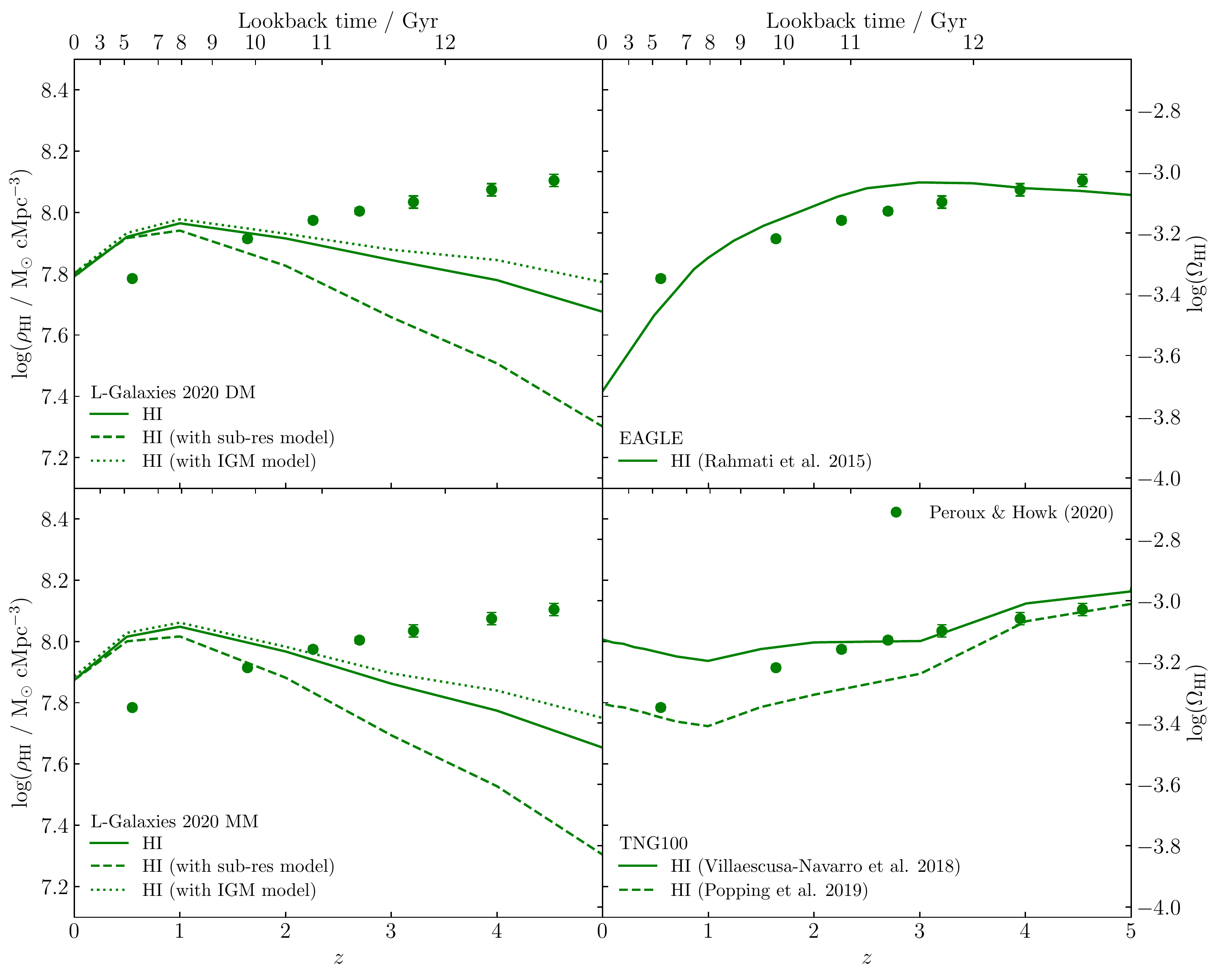} 
 \caption{Evolution in \textsc{Hi} mass density in galaxies from $z=5$ to 0. Each panel corresponds to a different galaxy evolution simulation. For the \lgaltt{} simulations (left panels), solid lines represent default results, dashed lines represent \HIDen{} when assuming a separate model for the \textsc{Hi} mass in sub-resolution systems, and dotted lines represent \HIDen{} when assuming a separate model for \textsc{Hi} mass in the IGM (see Section \ref{sec:Bary den evo}). For TNG100 (bottom right), the solid and dashed lines show complementary measurements of \HIDen{} from two different works (see legend). In all panels, points denote observational data from \citetalias{Peroux&Howk20}.}
 \label{fig:Baryon_density_evo_HI}
\end{figure*}

While the cosmic \SFRDen{} plays an important role in the intrinsic evolution of the metal density in neutral gas, measurements of $\rho\sub{HI}$ are equally important in its observational estimation via Eqn. \ref{eqn:metalDensity_obs}. At $z\leq{}0.4$, cosmic $\rho\sub{HI}$ is typically inferred via measurements of $M\sub{HI}$ from the 21-cm emission line. An \textsc{Hi} mass function is then formed and integrated between reasonable limits to obtain $\rho\sub{HI}$. At $0.4\lesssim{}z\lesssim{}5.0$, measurements of the Lyman-$\alpha$ absorption line in the spectra of background quasars are used to determine the \textsc{Hi} column density, $N\sub{HI}$. The column density distribution function (CDDF) can then be formed, from which $\rho\sub{HI}$ is derived in a similar way to that at lower redshifts. \citetalias{Peroux&Howk20} provide a linear fit to the redshift evolution of $\rho\sub{HI}$ obtained from a compilation of recent surveys utilising these two techniques, given by, 

\begin{equation}\label{eqn:baryDensity_obs}
\rho\sub{HI,obs}(z) = \rho\sub{crit,0}\,(4.6\error{0.2}\times{}10^{-4})\,(1+z)^{0.57\pm{}0.04} - (1/0.76)\ \ \ ,
\end{equation}

where the correction factor of $1/0.76$ accounts for helium. The green points in Fig. \ref{fig:Baryon_density_evo_HI} represent this fit, plotted at the discrete redshifts of the binned DLA data used in Sections \ref{sec:Met den evo} and \ref{sec:ZDLA evo}. The recent analysis of \citet{Heintz+21}, which utilises [\textsc{Cii}] emission line measurements and a [\textsc{Cii}] -- \textsc{Hi} conversion factor calibrated to GRB host galaxies, found a similar rate of increase in $\rho\sub{HI}$ with redshift above $z\sim{}2$, with this trend flattening somewhat at $z\sim{}5-6$.

For the \lgaltt{} models, the \textsc{Hi} density shown in Fig. \ref{fig:Baryon_density_evo_HI} is calculated from the total hydrogen mass in cold gas, $M\sub{H}$, as,

\begin{equation}\label{eqn:baryDensity_theo}
\rho\sub{HI,theo} = \,\frac{\sum{(1-f\sub{H2})\,M\sub{H}}}{V\sub{tot}}\ \ \ ,
\end{equation}

where $V\sub{tot}$ is again the volume of the simulation box, and $f\sub{H2}$ is the molecular hydrogen fraction as calculated following the \citet{McKee&Krumholz10} gas partition formalism (see \citealt{Henriques+20}, section 2.2.3). For EAGLE, $\rho\sub{HI}$ is taken directly from a previous analysis by \citet{Rahmati+15}. In that study, the equilibrium \textsc{Hi} fraction in each gas particle is calculated by balancing the collisional and UVB photoionization rates (as parameterised by \citealt{haardt2001}) with the Case A recombination rate, assuming a temperature of $10^{4}$ K for star-forming gas and self shielding in denser regions (see \citealt{rahmati13}). For TNG100, cosmic $\rho\sub{HI}$ is taken from two complementary analyses by \citet{villaescusa-navarro2018} and \citet{popping19}, which applied two different prescriptions for \textsc{Hi} -- H$_{2}$ gas partitioning. \citet{villaescusa-navarro2018} follow a similar prescription to \citet{Rahmati+15}, but adopt the UVB model of \citet{fg09}. \citet{popping19} instead first calculate the H$_{2}$ mass following the metallicity-dependent formalism of \citet{gnedin11}, then remove this from the total neutral gas budget to obtain the \textsc{Hi} mass.

We highlight here that these recipes also account for the presence of \textsc{Hi} in low-density, non-star-forming gas in EAGLE and TNG100. Such gas is not modelled in \lgaltt{}, contributing to the significantly lower \HIDen{} predicted at high redshift in that simulation, as discussed below.

The trend of increasing \HIDen{} with increasing redshift seen in observations is well reproduced by EAGLE and TNG100, as is shown in previous works (\eg{}\citealt{Rahmati+15,villaescusa-navarro2018,diemer19}). However, we note that the observed densities inferred from Lyman-$\alpha$ absorption at higher redshift directly probe only dense DLA gas with $N\sub{HI} \geq{} 10^{20.3}\,\tn{cm}^{-2}$, with an extrapolation down to lower column densities required (typically down to $N\sub{HI} \sim{} 10^{12}\,\tn{cm}^{-2}$, see \citealt{Zafar+13,berg19}). Less than 40 per cent of the total \textsc{Hi} mass is actually at DLA densities in TNG100, at all redshifts, and the evolution of this dense neutral gas is flatter than shown for all \textsc{Hi} gas in Fig. \ref{fig:Baryon_density_evo_HI} (see \citealt{Hassan+20}).

\lgaltt{} only reproduces the observed trend in \HIDen{} below $z\sim{}1$, albeit with a higher normalisation than observations (while still matching the observed \textsc{Hi} mass function from \citealt{Zwaan+05} at $z=0$, see \citealt{Yates+21a}). Above $z\sim{}2$, both \lgaltt{} models suggest a \textit{decrease} in $\rho\sub{HI}$ with redshift, in contrast to observations. This discrepancy has also been seen in earlier versions of \lgal{} \citep{Martindale+17}, as well as the \textsc{Galform}, \textsc{SantaCruz}, and GAEA semi-analytic models \citep{Lagos+11,Berry+14,Spinelli+20}, and the fiducial analytic model presented by \citet{Theuns21}. These works suggest that this \HIDen{} deficit is due to a lack of adequate accounting for \textsc{Hi} in sub-resolution haloes and/or neutral gas outside galaxies. The modelling of the ultraviolet background (UVB) radiation and of H$_{2}$ could also play a role, although we note that the uniform UVB models adopted by the simulations considered here predict a similar evolution in the \textsc{Hi} photoionisation rate with redshift (see \citealt{Faucher-Giguere20}).

When considering \textsc{Hi} in sub-resolution haloes, we note that \citet{Di_Gioia+20} utilised a halo occupation distribution (HOD) in GAEA to more accurately model the number and \textsc{Hi} content of such systems. We mimic this approach here for \lgaltt{}, by taking model \textsc{Hi} measurements only from resolved subhaloes with masses above $M_{200} = 10^{9.2}\Msun$ and adding a fixed contribution from lower-mass systems of $1.5\times{}10^{-5}\rho\sub{crit}\,\Msun/\tn{Mpc}^{3}$ at all redshifts (see \citealt{Di_Gioia+20}, fig. 6b). This evolution is shown by the dashed green lines in the left-hand panels of Fig. \ref{fig:Baryon_density_evo_HI}. As found by \citet{Spinelli+20} and \citet{Di_Gioia+20}, this approach still under-estimates the \textsc{Hi} densities observed at $z\gtrsim{}3$. Indeed, the HOD-based prescription actually predicts lower cosmic \textsc{Hi} densities at high redshift than are obtained when including the \textsc{Hi} masses from sub-resolution haloes in \lgaltt{} (solid green lines). \citet{Di_Gioia+20} found that an \textit{a posteriori} increase in the \textsc{Hi} content of \textit{all} galaxies by a factor of two was also required before \HIDen{} was significantly increased in GAEA. This suggests that corrections to the gas content of sub-resolution systems alone is not enough to resolve the discrepancy in \textsc{Hi} density seen between semi-analytic models and observations.

When considering \textsc{Hi} outside of galaxies, we note that \citet{villaescusa-navarro2018} find $\sim{}6$ per cent of the cosmic \textsc{Hi} at $z=3$ in TNG100 lies beyond the subhalo boundary, with this percentage increasing to 20 per cent by $z=5$. Such a neutral intergalactic medium (IGM) is not modelled at all in semi-analytic models (see also \citealt{lagos2014,Xie+17}). Therefore, we mimic the effect of including this IGM \textsc{Hi} gas, by adding its fraction as seen in TNG100 (see \citealt{villaescusa-navarro2018}, fig. 3) to that found inside galaxies in \lgaltt{}. This adapted \HIDen{} evolution is shown by the dotted green lines in the left-hand panels of Fig. \ref{fig:Baryon_density_evo_HI}. While this correction does increase \HIDen{} in \lgaltt{}, especially at higher redshift, it still falls short of that observed at $z\gtrsim{}3$. This indicates that larger \textsc{Hi} masses \textit{within resolved subhaloes} are still required in semi-analytic models at high redshift.

This missing neutral gas should presumably lie in the outskirts of galaxies and/or in the inner CGM. We find that a substantial amount ($\sim$ 65 per cent) of the \textsc{Hi} gas within subhaloes in TNG100 lies beyond $2r\sub{e}$ at $z\gtrsim{}3$, where $r\sub{e}$ is the stellar half-mass radius \citep[see][for observable implications]{Byrohl+21}. A similar conclusion has been drawn for the EAGLE simulation by \citet{Garratt-Smithson+21}, who find up to 50 per cent of \textsc{Hi} gas within massive subhaloes is in the CGM at $z=2$. Likewise, \citet{stevens19} point to the importance of the CGM in TNG100 when comparing \textsc{Hi} masses to wide-aperture observations of the 21cm line at lower redshift.

Given this, we make a simple test to see if an \textit{ad hoc} correction to the \HIDen{} within subhaloes in \lgaltt{} could also resolve the discrepancy in \MetDenNeu{} at high $z$ in Fig. \ref{fig:Metal_density_evo}. We do this by assuming that the same deficit in \textsc{Hi} seen between \lgaltt{} and observations in Fig. \ref{fig:Baryon_density_evo_HI} is also present for metals, such that the mean cosmic cold gas metallicity in the simulation is unchanged. The adapted \MetDenNeu{} evolution that this simple test produces is shown as green dotted lines in the left-hand panels of Fig. \ref{fig:Metal_density_evo}.

At $z\lesssim{}2$, the lower \MetDenNeu{} caused by this correction (due to the apparent \textit{excess} of \textsc{Hi} in the models) brings both the DM and MM into improved agreement with DLA observations. At $z\gtrsim{}2$, this \textit{ad hoc} correction also nicely improves agreement for \lgaltt{} DM. However, this version of the simulation is already known to over-estimate ISM metallicities in high-$z$ galaxies (see \citealt{Yates+21a}). For \lgaltt{} MM, which reproduces ISM metallicities in high-$z$ galaxies well (see Section \ref{sec:ZDLA evo}), \MetDen{} remains too low compared to DLA observations when assuming an unchanged mean cosmic cold gas metallicity. Instead, we find that a mean metallicity for the added gas of $12+\tn{log(O/H)}\sim{}8.45$ is required at $z\sim{}4$ to fully resolve the \MetDenNeu{} discrepancy seen, which is roughly the metallicity of galaxies with $\logMm{} = 10.5$ at that redshift in \lgaltt{} MM.

Such high-metallicity material would require a significant increase in \SFRDen{}, and consequently $\rho_{*}$, at high redshift, which may bring \lgaltt{} into better agreement with TNG100 (although worse agreement with observations from \citealt{Madau&Dickinson14}). However, this outcome is far from certain, as an increase in cold gas mass would also necessitate changes to the rates of gas cooling, H$_{2}$ formation, and other processes in \lgaltt{}. Clearly, further study of the implications (and physical motivations) for such a change in semi-analytic models is required before a more definitive statement can be made. Such an \textsc{Hi}-based correction would also do little to resolve the smaller deficit in \MetDen{} seen in EAGLE, which reproduces \HIDen{} reasonably well at high redshift.

We therefore conclude that a simple increase in the amount of neutral gas in or around galaxies in simulations is unlikely to provide a complete solution to the neutral gas metal discrepancy seen in Fig. \ref{fig:Metal_density_evo}. However, it is likely to be a contributing factor in the case of semi-analytic models, in combination with the other factors discussed in Section \ref{sec:ZDLA evo}.

\subsection{Neutral gas metallicity evolution}\label{sec:ZDLA evo}

In addition to the neutral gas density discussed in Section \ref{sec:Bary den evo}, the mean neutral gas metallicity, \MetNeu{}, also plays an important role in determining \MetDenNeu{} in observations via Eqn. \ref{eqn:metalDensity_obs}. Fig. \ref{fig:Metal_density_evo} shows that the evolution in \MetDenNeu{} as probed by DLAs is flatter than that seen in galaxy evolution simulations (and in observational $\rho\sub{met,tot}$ estimates, see also \citealt{rafelski2014}). However, the corresponding evolution in \MetNeu{} for DLAs is actually \textit{steeper} than seen in simulations and other ISM observations \textit{at fixed mass}. For example, \citet{De_Cia+18} find an evolution in \MetNeu{} of $\sim{}0.63$ dex from $z\sim{}3.2$ to 0.5 for their DLA sample. In contrast, estimates of the mean ISM oxygen abundance evolution at fixed mass from recent emission-line studies (using direct and indirect metallicity diagnostics) do not exceed $\sim{}0.38$ dex across the same redshift interval (\eg{}\citealt{Izotov+15,Hunt+16,Sanders+16,Sanders+21}). Studies which also utilise GRB absorption-line metallicities at high redshift (which, like DLAs, should be unbiased in terms of host galaxy luminosity) also find similar results (\eg{}\citealt{Thoene+13,Graham+19,Yates+21a}). In several recent cosmological simulations, the evolution in the ISM oxygen abundance of star-forming galaxies at fixed mass is also found to be only $0.25-0.45$ dex from $z\sim{}3$ to 0 (see \citealt{Yates+21a} section 4.2, and references therein).

\begin{figure*}
\centering
 \includegraphics[angle=0,width=0.99\linewidth]{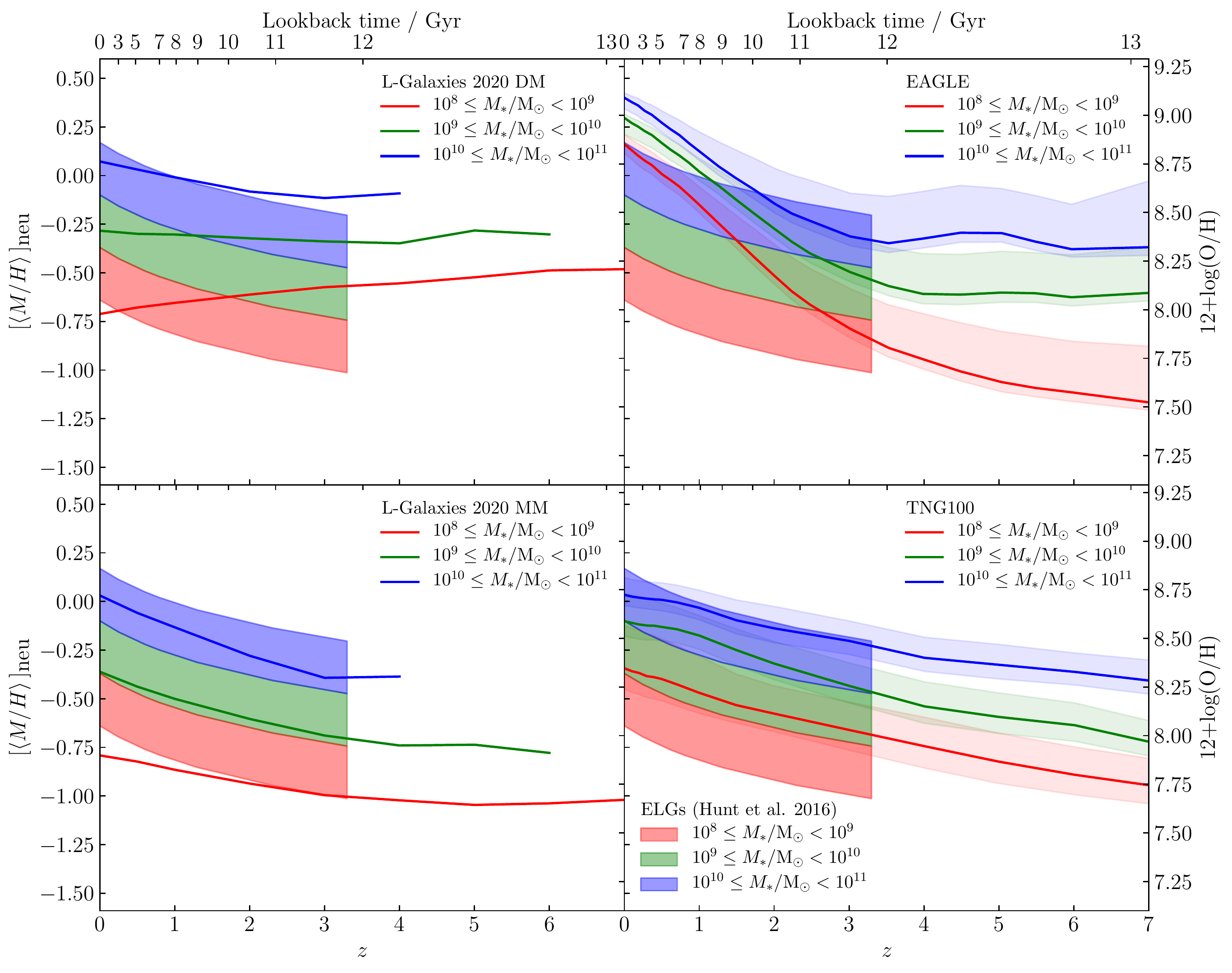}
 \caption{Evolution of gas-phase metallicity in three fixed-mass bins (see legend). Observational results for ELGs from $z=0$ to 3.3 by \citet{Hunt+16} are shown as dark shaded regions. The mean, \textsc{Hi}-weighted gas metallicities for model galaxies in the four simulations considered here are also shown. Light shaded regions represent the uncertainty in the DLA density threshold for EAGLE and TNG100 (see Section \ref{sec:Measuring densities}). All metallicities are normalised to the solar photospheric value of $M\sub{met,\astrosun}/M\sub{H,\astrosun} = 0.0181$ from \citet{Asplund+09}.}
 \label{fig:MH_evo_fixedMassBins}
\end{figure*}

One possible explanation for this discrepancy is that the average mass of DLA host galaxies evolves with time. This may seem unlikely, as one of the main strengths of DLAs is that their detectability is not biased towards brighter (\ie{}more massive) systems at any particular redshift. Typical optical and UV sensitivities are also well below that required to detect the metal absorption lines seen in current DLA samples (see \citetalias{Peroux&Howk20}, section 3.4). Nonetheless, the masses of DLA host galaxies are currently poorly constrained. For example, only $\sim{}5$ of the 235 DLAs in the \citet{De_Cia+18} sample have identified galaxy counterparts with measured stellar masses and impact parameters \citep{krogager2013,Fynbo+13,Christensen+14,Augustin+18,Rhodin+21}.

Therefore, in this section we attempt to quantify a possible mass dependence of \MetNeu{} measurements from a theoretical perspective. To do this, metallicities from the simulations are measured in an equivalent way to the DLA observations, by taking the log of the \textsc{Hi}-weighted mean metal-to-hydrogen mass ratio in cold, dense gas,
\begin{equation}\label{eqn:MeanMH}
    [\langle{}\tn{M/H}\rangle{}]\sub{neu} = \logten{}\left[\frac{\sum\left(M\sub{met}/M\sub{H}\right)\,M\sub{HI}}{\sum{M\sub{HI}}}\right] - \logten{}\left[\frac{M\sub{met,\astrosun}}{M\sub{H,\astrosun}}\right]\ \ \ ,
\end{equation}
with both model and observational values normalised to the solar photospheric value of $Z\sub{\astrosun}/X\sub{\astrosun} = M\sub{met,\astrosun}/M\sub{H,\astrosun} = 0.0181$ from \citet{Asplund+09}. The total metal and hydrogen masses in the cold, dense gas of each model galaxy are used, such that $[\langle{}\tn{M/H}\rangle{}]\sub{neu}$ represents the cosmic average metallicity for galaxies. Weighting each galaxy's contribution by their \textsc{Hi} mass mitigates somewhat any differences in the fraction of star-forming galaxies among the various observational and model samples.

\subsubsection{Metallicity evolution at fixed mass}\label{sec:Metallicity evolution at fixed mass}
Before comparing directly with DLA metallicities, we first show a comparison between simulations and the expected ISM metallicity evolution in star-forming emission-line galaxies (ELGs) as probed by composite spectra dominated by \textsc{Hii} regions. Fig. \ref{fig:MH_evo_fixedMassBins} shows the evolution in \MetNeu{} from \lgaltt{}, EAGLE, and TNG100, as defined by Eqn. \ref{eqn:MeanMH}, for three, 1-dex-wide, fixed-mass bins (solid lines), alongside the evolution from a compilation of ELG samples by \citet{Hunt+16} (dark shaded regions). These observations were re-calibrated to the \citet{Pettini&Pagel04} N2 strong-line diagnostic, which returns good agreement with various direct metallicity measurements at both low and high redshift \citep{Andrews&Martini13,Yates+21a}. However, we acknowledge that other strong-line diagnostics predict higher metallicities at low redshift (\eg{}\citealt{Maiolino+08,Zahid+14}).

The evolution in model \textsc{Hi}-weighted neutral gas metallicities shown here is similar to that previously reported when using SFR-weighted methods and oxygen abundances in each of the simulations considered (see \citealt{DeRossi+17} for EAGLE, \citealt{Torrey+19} for TNG100, and \citealt{Yates+21a} for \lgaltt{}). Although, we note that plotting the average \MetNeu{} in 1-dex-wide mass bins here naturally leads to a slightly steeper evolution than would be seen for an exact fixed mass, due to the increasing contribution to cosmic metals (as well as cosmic \textsc{Hi} and SFR, see \citealt{lagos2014,genel14,Spinelli+20}) from higher-mass, more metal-rich galaxies over cosmic time, at least for systems below $\logMm\sim{}10.5$.

Fig. \ref{fig:MH_evo_fixedMassBins} shows that only \lgaltt{} MM is able to reproduce the fixed-mass metallicity evolution for ELGs observed by \citet{Hunt+16} back to $z\sim{}3.3$ (with the exception of the lowest-mass galaxies at the lowest redshifts). \lgaltt{} DM returns no clear evolution in \MetNeu{} at all, due to an over-retention of newly-synthesised metals in the ISM at early times (see \citealt{Yates+21a}). TNG100 returns a higher normalisation than observed by \citet{Hunt+16}, and EAGLE has both a higher normalisation and steeper slope. Although, we note that the Recal-L025N0752 version of EAGLE returns lower metallicities in low-mass galaxies than the Ref-L100N1504 version used here, due to its stronger SN feedback \citep{schaye15,DeRossi+17}. We also find that, at high redshift, gas with $n\sub{gas} > 10^{-1.0}\,\tn{cm}^{-3}$ in EAGLE (\ie{}the upper limit of the uncertainty bands) appears to have much higher metallicity, by $\gtrsim{}0.2$ dex, than gas with $n\sub{gas} > 10^{-1.3}\,\tn{cm}^{-3}$ (solid lines), indicating a strong dependence of metallicity on gas density. This has also been identified in EAGLE by \citet{DeRossi+17} and \citet{Garratt-Smithson+21}.

In general, low-mass galaxies are much more metal rich by $z=0$ in EAGLE and TNG100 than in \lgaltt{}. The Ref-L100N1504 EAGLE simulation even predicts super-solar ISM metallicities in galaxies with masses as low as $\logMm{}\sim{}8$ at $z=0$. This is predominantly due to differences in the SN feedback efficiency in these simulations. As discussed in Section \ref{sec:sim differences}, the reference EAGLE simulation shown here has relatively low SN feedback efficiency, leading to higher metallicities in lower-mass systems (\citealt{schaye15,DeRossi+17}). This provides reasonable agreement with older, indirect ISM metallicity measurements for ELGs at low redshift (\eg{}\citealt{Tremonti+04,Kobulnicky&Kewley04}). On the other hand, the increased metal ejection efficiency from SNe in \lgaltt{} MM leads to better agreement with more recent indirect and direct metallicity measurements (\eg{}\citealt{Hunt+16,Dopita+16,Yates+20}).

Information on the gas-phase mass -- metallicity relation (MZ$\sub{g}$R) at various redshifts is also obtainable from Fig.~\ref{fig:MH_evo_fixedMassBins}. For example, at $z\sim{}3-4$, we can see that the MZ$\sub{g}$R in \lgal{} MM is steeper than that in EAGLE (\ie{}the separation between the fixed-mass lines in Fig.~\ref{fig:MH_evo_fixedMassBins} is greater), which in turn is steeper than that in TNG100. This ordering is reflective of the relative \MetDenNeu{} in these simulations at these redshifts, as shown in Fig.~\ref{fig:Metal_density_evo}, demonstrating the importance of the metal content in lower-mass galaxies. We further note that, for gas with $n\sub{gas} > 10^{-1.0}\,\tn{cm}^{-3}$, the EAGLE MZ$\sub{g}$R becomes as flat as that in TNG100 at $z\sim{}3-4$, due to the significant presence of metals in the densest regions.

\begin{figure*}
\centering
 \includegraphics[angle=0,width=0.99\linewidth]{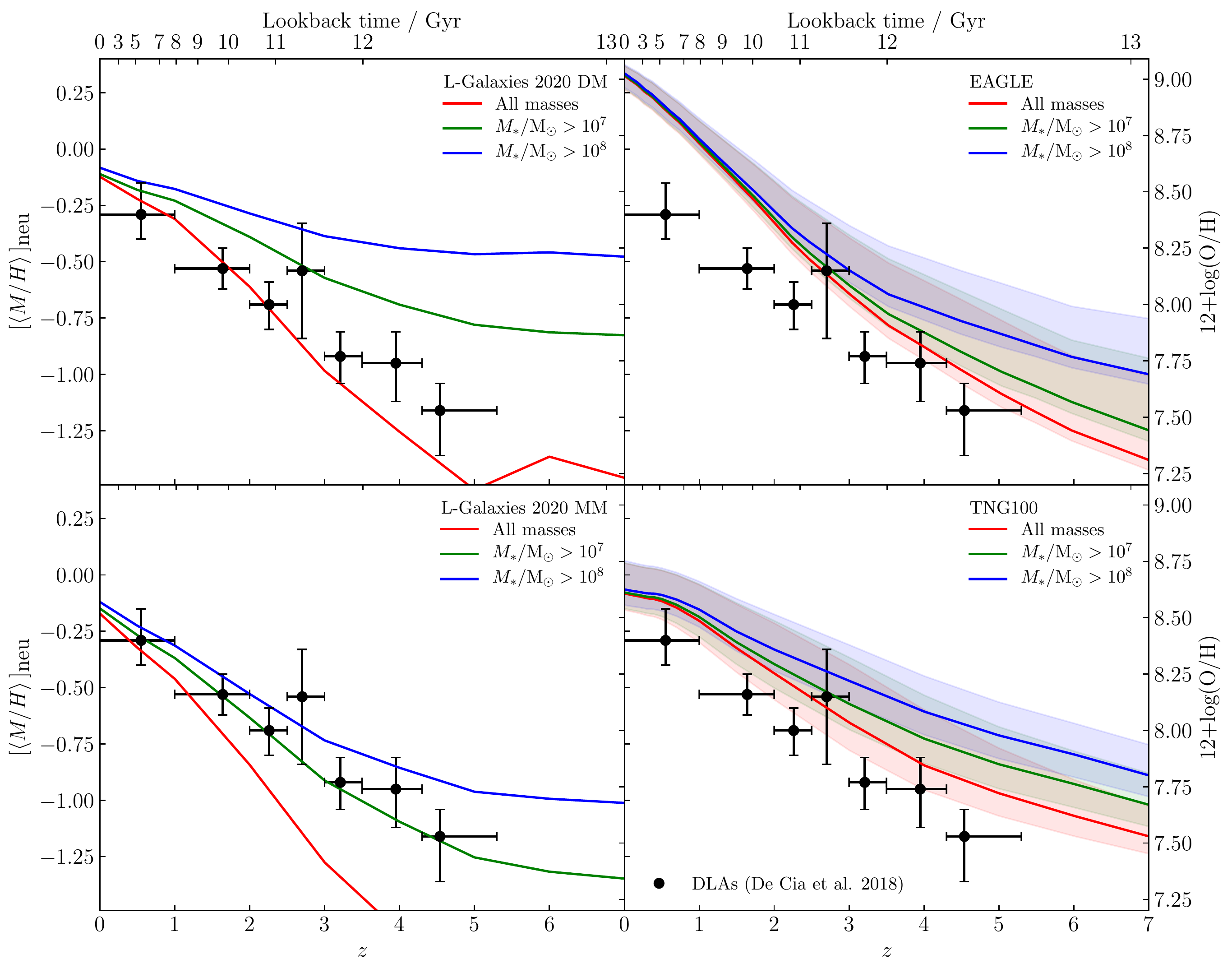}
 \caption{Evolution of the mean, \textsc{Hi}-weighted gas metallicity for DLAs (\citealt{De_Cia+18}, black points) and the four simulations considered here. Solid lines represent the \MetNeu{} evolution for model galaxies above a certain mass threshold (see legend). Light shaded regions represent the uncertainty in the DLA density threshold for EAGLE and TNG100 (see Section \ref{sec:Measuring densities}). All metallicities are normalised to the solar photospheric value of $M\sub{met,\astrosun}/M\sub{H,\astrosun} = 0.0181$ from \citet{Asplund+09}.}
 \label{fig:MH_evo_lowerMassThreshold}
\end{figure*}

\subsubsection{Mass dependencies in DLA host samples}\label{sec:M/H for DLAs}
Fig. \ref{fig:MH_evo_lowerMassThreshold} shows the observed evolution in \textsc{Hi}-weighted \MetNeu{} for the \citet{De_Cia+18} DLA sample (black circles) from $z\sim{}5$ to 0. Solid lines denote the corresponding evolution for galaxies in the four simulations considered here, for three fixed lower-mass thresholds (see legend). A clear evolution in \MetNeu{} is seen, in contrast to earlier determinations using smaller samples and relatively undepleted metallicity indicators such as zinc (\eg{}\citealt{Vladilo+00}). 

When considering galaxies without any lower-mass threshold (red lines), the \MetNeu{} evolution seen in simulations is indeed steeper than seen at fixed mass in Fig. \ref{fig:MH_evo_fixedMassBins}. This is due to the combination of the galaxy mass -- metallicity relation and the natural increase over cosmic time of the mean mass of volume-limited galaxy samples in a $\Lambda$CDM cosmology.

In \lgaltt{} DM, this evolution for all galaxies is close to that observed for DLAs, suggesting that current DLA samples might not have a significant mass bias compared to the overall galaxy population. However, we note again that \lgaltt{} DM produces overly-enriched galaxies at high redshift and is therefore disfavoured for this particular analysis. In \lgaltt{} MM, the evolution of \MetNeu{} for all galaxies is \textit{steeper} than observed for DLAs. Given that this simulation matches the observed metallicity evolution in ELGs at fixed mass (see Section \ref{sec:Metallicity evolution at fixed mass}), this suggests a shift in the mass distribution of DLA hosts with respect to the overall galaxy population at high redshift.

Dust corrections could also play a role, as these tend to be greater for metal-rich (\ie{}lower-redshift) systems, thus influencing the slope of the observed \MetNeu{} evolution for DLAs (see \citealt{Krogager+19}). However, the fact that the normalisation of \MetNeu{} in \lgaltt{} MM is in good agreement with DLA observations at $z\lesssim{}1$ suggests that over-estimated dust depletion corrections in metal-rich systems are not a main factor here.

The most straightforward way to model a shift in the mass distribution of galaxies at high redshift in simulations is to impose a redshift-invariant lower-mass limit (see also \citealt{dvorkin2015}). When doing this, Fig. \ref{fig:MH_evo_lowerMassThreshold} shows that the observed \MetNeu{} evolution from DLAs is best reproduced in \lgaltt{} MM when only galaxies with masses above $\logMm{}\sim{}7.0-7.5$ are considered.

We note here that galaxies of mass below $\logMm{}\sim{}7.0$ live in DM subhalos with $M_{200}\sim{}4\times{}10^{7}\,h^{-1}\,\Msun$ in the simulations considered here, comprising only $\sim{}20$ DM particles. These systems are therefore at the resolution limit, and their baryonic properties are unlikely to be modelled accurately. Above $\logMm{}\sim{}8.0$, galaxies are hosted by subhaloes with $M_{200} \gtrsim{} 1.3\times{}10^{9}\,h^{-1}\,\Msun$, containing $\gtrsim 100$ DM particles and are therefore substantially better resolved. The lower-mass limit for DLA host galaxies preferred by \lgaltt{} MM is therefore at the boundary between well and poorly-resolved galaxies in these simulations. Its value is therefore open to some uncertainty and we have not attempted to determine it more precisely here.

Nonetheless, a lower-mass limit of $\logMm\sim{}7.5$ is consistent with recent determinations of DLA host masses in absorption-selected galaxy samples at $z=2-3$ (\eg{}\citealt{Krogager+20a,Rhodin+21}), although we note that such samples are biased towards metal-rich DLAs and therefore more massive systems. The lower limit determined from \lgaltt{} MM is also in good agreement with the approximate stellar mass estimates obtained using the metallicity- and redshift-dependent relation of \citet{Moller+13}. When assuming a conversion factor between emission-line-based and absorption-line-based metallicities of $C=0.44\error{0.10}$ from \citet{Christensen+14}, we find that 97 per cent of the DLA hosts in the \citet{De_Cia+18} sample are expected to have stellar masses above $\logMm=7.5$. The presence of a non-zero lower-mass limit and an increase in average mass over cosmic time is also consistent with the cross-correlation of DLAs and the Ly$\alpha$ forest \citep{Perez-Rafols+18a}.

In \lgaltt{} MM, a lower-mass limit of $\logMm=7.5$ returns an evolution in the \textit{mean} stellar mass of galaxies from $z=5$ to 0 of 1.26 dex, increasing from $\logMm = 8.13$ to 9.38. Due to the shape of the galaxy stellar mass function, the equivalent evolution in the \textit{median} stellar mass is only 0.18 dex, increasing from $\logMm = 7.76$ to 7.94. The mean virial mass of subhaloes increases by 0.61 dex from $\tn{log}(M_{200}/\Msun) = 10.57$ to 11.18 over the same period.

At $z=3$, the mean virial mass of $\tn{log}(M_{200}/\Msun) = 10.74$ we find is in good agreement with the estimate of 10.78 determined by \citet{Krogager+20a} from their DLA host galaxy modelling. Their model considers host galaxies down to $\logMm{}\sim{}6.5$ at $z=2$, which is an order of magnitude lower than the limit preferred by \lgaltt{} MM. The good correspondence between the two models therefore further suggests that the lowest-mass systems do not contribute significantly to DLA samples, perhaps due to their lower gas cross-sections. Indeed, we find that at $3\lesssim{}z\lesssim{}5$, where the discrepancy between observational and model \MetDenNeu{} estimates is greatest, galaxies with $\logMm{}\gtrsim{}7.5$ strongly dominate the \MetDenNeu{} budget in all four simulations, and sub-DLAs are expected to have only a minor contribution (\eg{}\citealt{Berg+21}). Our findings are also consistent with those of \citet{Erkal+12}, who used a high-resolution hydrodynamical simulation to determine that hosts with $\tn{log}(M_{200}/\Msun) < 11$ provide the dominant contribution to DLA cross sections at all column densities at $z=3$.

Importantly, a lower-mass limit of $\logMm\sim{}7.5$ also brings \lgaltt{} MM into better agreement with the observed \MetDenNeu{} from DLAs in Fig. \ref{fig:Metal_density_evo} (dashed green line). However, an additional discrepancy still remains. This is likely due to the residual deficit in neutral gas present in subhaloes at high redshift in \lgaltt{}, as discussed in Section \ref{sec:Bary den evo}.

EAGLE and TNG100 both return higher metallicities than seen for DLAs at all redshifts, regardless of the lower-mass threshold assumed. This is particularly true for denser gas in EAGLE (upper limit of the uncertainty bands), and cannot be explained by differences among strong-emission-line metallicity diagnostics. This suggests that the higher \MetDenNeu{} at high redshift in EAGLE (see Fig. \ref{fig:Metal_density_evo}) could be partly due to an over-retention of metals in the dense ISM, at the expense of the partially-ionised gas phases. Whereas the higher \MetDenNeu{} seen in TNG100 may be due to an over-production of metals altogether (see Section \ref{sec:SFRD evo}).

\subsubsection{Other factors affecting $[\langle{}M/H\rangle{}]$ evolution} 
In addition to possible mass dependencies affecting the observed DLA metallicity evolution, a systematic evolution in ISM radial metallicity gradients could also play a role. However, observations of such gradients in galaxies back to $z\sim{}2.5$ suggest there is little to no overall evolution, with star-forming galaxies typically having slightly negative gradients at all redshifts (\eg{}\citealt{Stott+14,Leethochawalit+16,Patricio+19,Curti+20b,Gillman+21}). A similarly mild evolution in ISM metallicity gradients at fixed mass in galaxy discs is found in galaxy evolution simulations (\eg{}\citealt{Gibson+13,Ma+17,Yates+21a}), although \citet{Hemler+20} find a somewhat stronger evolution in TNG50.

A systematic evolution in typical DLA impact parameters, $b$, could also affect the inferred \MetNeu{} evolution. This could be the case even without an evolution in typical ISM metallicity gradients, if DLAs also probe dense gas outside of galaxies. Indeed, \citet{Christensen+14} find a predicted decrease in the metallicity from the centre of host galaxies to the DLA itself of -0.022\error{0.004} dex/kpc, with \citet{Krogager+20a} similarly finding -0.019\error{0.008} dex/kpc for $L^{*}$ galaxies. However, the offset this implies between \MetNeu{} measurements from DLAs and model galaxies is mitigated somewhat by the fact that the mean values calculated at each redshift are \textsc{Hi} weighted, reducing the impact of DLAs with large $b$, as these systems tend to have lower \textsc{Hi} column densities (see \eg{}\citealt{krogager2017}). In addition, we find that those few DLAs in the \citet{De_Cia+18} sample which do have a large measured $b$ of $\gtrsim{}12$ kpc \citep{Christensen+14,Augustin+18} all have \textit{higher} metallicities than the average for DLAs at their redshift. Cold, dense gas clumps observed in galactic outflows are expected to be relatively metal rich \citep{Peroux+20}, although dense accreting gas should have lower metallicities (\eg{}\citealt{Garratt-Smithson+21}).

Finally, systematic differences between emission-based and absorption-based metallicity measurements could also affect the comparisons made between various ISM metallicity measurements in this section. As mentioned in Section \ref{sec:Obs}, emission-based abundances from QSO-DLA host galaxies can differ from the absorption-based measurements by up to $\sim{}0.6$ dex \citep{peroux2014,rahmani2016}. However, this difference includes an intrinsic contribution from any metallicity gradient present, as the QSO-DLAs used in those studies were offset from the central emission-line regions by $10-30$ kpc. Systems with smaller impact parameters, such as SBS 1543+593 with $b\sim{}3$ kpc, instead return differences between absorption-based and emission-based metallicities of $\lesssim{} 0.1$ dex \citep{schulte-ladbeck2004,schulte-ladbeck2005,bowen2005}. Likewise, measurements for the host galaxy of GRB121024A, where the emission- and absorption-line regions are expected to be roughly co-spatial, suggest a discrepancy of only $\sim{}0.3$ dex \citep{Friis+15}. Studies of the blue compact dwarf galaxy, I Zw 18, which compare absorption-based abundances from its \textsc{Hi} regions with \textit{direct} emission-based abundances from electron temperature measurements, also return a discrepancy of only $\sim{}0.17$ dex \citep{aloisi2003,Lecavelier_des_Etangs+04,Pequignot08,lebouteiller2013}. Nonetheless, further investigation, particularly at higher redshifts, is required to better determine any systematic differences between emission-line and absorption-line techniques.

In conclusion, we find that the mean \textsc{Hi}-weighted metallicity in galaxies at high redshift in the \lgaltt{}, EAGLE, and TNG100 simulations is sensitive to the minimum stellar mass of the population. Simulations can therefore be used in conjunction with the currently-observed \MetNeu{} evolution to constrain the mass distribution of DLA host galaxies at high redshift. A lower-mass limit of $\logMm{}\sim{}7.5$ would reconcile the \lgaltt{} MM with both observed DLA metallicities and \MetDenNeu{}. However, such an assumption does not improve the correspondence between observations and EAGLE or TNG100. These simulations over-predict DLA metallicities, possible due to an over-retention of metals in very dense gas in EAGLE, and an over-production of metals in TNG100. Other biases in observational samples could also play a role, and future studies of DLA host galaxies will be invaluable for improving constraints (see Section \ref{sec:future_prospects}).


\section{Future Prospects}\label{sec:future_prospects}

As mentioned in Section \ref{sec:Intro}, the dominant contributors to the total metal budget, other than DLAs, have not been measured at higher redshifts. The hot phase in particular, which is mainly traceable in absorption through \eg{}\ion{Ne}{viii} in the UV \citep{frank2018, burchett2019} and \ion{O}{vii} and \ion{O}{viii} in the X-ray \citep{nelson18b}, are presently limited to $z\lesssim{}1.4$ \citep{ettori2015}. Alternatives using novel techniques or metal transitions
\citep[e.g.][]{pettini1986,anderson2016, anderson2018, Fresco+20} will likely play a role. Ultimately, more robust measurements will have to await the next generation of X-ray facilities such as Athena \citep{Kirpal13} and Lynx \citep{lynx2018}.

At intermediate redshift, the census of metals in the Universe is still open. Ambitious surveys of order a million quasar spectra will provide a statistical sample of absorbers sufficient to obtain a definite picture of the metal content in baryons in the Universe over its entire lookback time. These efforts will rely on the availability of the next generation of multi-object spectrographs, \eg{}4MOST \citep{deJong19}, WEAVE \citep{Dalton12}, and MSE \citep{MSE19}. Similarly, 21 cm measurements beyond $z\sim{}1$ from \eg{}the MeerKAT LADUMA survey \citep{Blyth+16} will help determine to what extent, if at all, DLAs provide biased \HIDen{} estimates. 

A more precise census of the star formation occurring in lower-mass galaxies at high redshift will also only be possible with the enhanced capabilities of JWST \citep{Gardner06}. Instruments such as NIRCam and MIRI will allow SFRs to be measured for many faint galaxies at $z\sim{}2.5$ and beyond, as well as providing additional photometry in the rest-frame NIR for galaxies at $z\gtrsim{}4$, potentially helping to constrain stellar masses for high-redshift DLA hosts.

The unparalleled power of JWST/NIRSpec and VLT/BlueMUSE \citep{Richard19} will also enable a large number of precise gas-phase metallicity measurements in and around galaxies to be made at high redshift. These, along with detailed high-redshift metallicity calibrations from VLT/MOONS \citep{Cirasuolo20}, will enable a much more detailed comparison between the metal density in DLA samples and that of the general star-forming galaxy population.

Finally, we highlight that the present study is concerned primarily with the cosmic metal density in the Universe, rather than the metal content of DLAs specifically. The ideal way to study the latter in simulations would be to produce mock line-of-sight spectra and modelling a range of metal absorption lines from multiple metal ions such as Si, S, and Fe. Assuming the same parameters as in DLA observations (\eg{}pixel resolution and signal-to-noise ratio) would also be important, because the column density of pixels along the line of sight and the damping wings are both needed to identify DLAs and compute their properties realistically. Some works have already adopted methods similar to this for selecting and analysing DLAs in simulations (\eg{}\citealt{pontzen2008,bird14,Berry+16,Di_Gioia+20,Hassan+20,Marra+21}), and further effort in this direction would be invaluable as the focus of future work with galaxy evolution simulations.


\section{Conclusions}\label{sec:Conclusions}

In this work, we compare recent observational results on the cosmic density of metals, star formation, neutral gas, and metallicity with that found in four cosmological galaxy evolution simulations: EAGLE, TNG100, and two versions of \lgaltt{}. This analysis allows us to investigate under what conditions the high \MetDenNeu{} estimates observed in high-redshift DLAs are possible from a theoretical perspective. Our main findings are as follows:

\begin{itemize}
\item The fraction of all metals that are in neutral gas at $z\gtrsim{}3$ is lower in models than inferred from DLAs. Observations of DLAs at $z\sim{}4$ suggest that \MetDenNeu{} constitutes $\sim{}87$ per cent of the total expected metal budget, when assuming the \SFRDen{} from \citet{Madau&Dickinson14}. In contrast, EAGLE, TNG100, and \lgaltt{} all suggest $<40$ per cent at the same redshift (see Section \ref{sec:Met den evo}).
\item In EAGLE and both versions of \lgaltt{}, the majority of metals at all redshifts are found in hot, low-density gas, despite the range of feedback efficiencies used in these simulations. This phase is particularly difficult to observe at high redshift, and plays an important role in accounting for the `missing metals problem'. In TNG100, the dominant contribution at $3\lesssim{}z\lesssim{}5$ is shared roughly equally between this low-density gas component, stars, and neutral gas (see Section \ref{sec:Met den evo}).
\item The absolute value of \MetDenNeu{} measured from DLAs at high redshift is well reproduced by TNG100. However, in order to achieve this, a  larger \SFRDen{} than is observed from FUV observations is needed. This tension highlights a possible incompatibility between the currently observed \MetDenNeu{} and \SFRDen{} (see Section \ref{sec:SFRD evo}).
\item The observed \HIDen{} at $z\gtrsim{}3$ is significantly under-estimated by \lgaltt{}, as has been seen in other semi-analytic models. This gas is most likely to be missing from the outskirts of resolved galaxies or the inner CGM, rather than from sub-resolution systems or the IGM. Accounting for this missing ISM material could help improve the correspondence between \lgaltt{} and the observed \MetDenNeu{} from DLAs (see Section \ref{sec:Bary den evo}).
\item The discrepancy in \MetDenNeu{} between simulations and observations at high redshift could also be resolved if the mass distribution of DLA host galaxies is shifted to higher masses than the overall galaxy population. For example, when assuming a fixed lower-mass threshold of $\logMm{}\sim{}7.5$, the observed evolution in \MetDenNeu{}, \SFRDen{}, and \MetNeu{} are simultaneously reproduced by \lgaltt{} MM (see Section \ref{sec:ZDLA evo}).
\item Both EAGLE and TNG100 appear to over-predict ISM metallicities for ELGs and DLAs at all redshifts. This could be partly due to an over-retention of metals in the densest gas in EAGLE, and an over-production of metals altogether in TNG100 (see Section \ref{sec:ZDLA evo}), although comparison to the redshift evolution of gas-phase metallicities is complicated by several systematic uncertainties.
\end{itemize}

We conclude that considering the simultaneous evolution of \MetDenNeu{}, \SFRDen{}, \HIDen{}, and \MetNeu{} is crucial when trying to constrain the key physical processes driving galaxy evolution. Future observations of the metal content in hot, diffuse gas at high redshift, and of the host galaxy properties of DLAs, are also required to help break the degeneracies revealed between these key cosmic properties by cosmological galaxy formation simulations.

\section*{Acknowledgements}
The authors would like to thank the referee for their constructive and insightful comments, as well as Lise Christensen, Payel Das, Denis Erkal, Paolo Molaro, and Marta Spinelli for valuable discussions during the undertaking of this work. RMY acknowledges funding from the UK Research and Innovation council (grant number MR/S032223/1). CP thanks the Alexander von Humboldt Foundation for the granting of a Bessel Research Award held at MPA where this work was initiated. DN acknowledges funding from the Deutsche Forschungsgemeinschaft (DFG) through an Emmy Noether Research Group (grant number NE 2441/1-1). Part of this research was carried out using the High Performance Computing resources at the Max Planck Computing and Data Facility (MPCDF) in Garching, operated by the Max Planck Society (MPG).

\section*{Data Availability}
Data directly related to this publication and its figures is available at \href{www.tng-project.org/yates21}{www.tng-project.org/yates21} and \href{robyatesastro.wixsite.com}{robyatesastro.wixsite.com}. The \lgaltt{} source code, as well as example output catalogues from the default and modified models, are publically available at \href{https://lgalaxiespublicrelease.github.io/}{lgalaxiespublicrelease.github.io/}. Complete output catalogues for the \lgaltt{} default model are also available on the \textsc{Millennium} database \citep{Lemson+06} at \href{http://gavo.mpa-garching.mpg.de/Millennium/}{gavo.mpa-garching.mpg.de/Millennium/}. The IllustrisTNG simulations are publicly available and accessible in their entirety at \href{www.tng-project.org/data}{www.tng-project.org/data} \citep{nelson19a}. The EAGLE simulation is publicly available \cite[see][for the original data release description]{McAlpine+16}.


\bibliographystyle{mnras}
\bibliography{robyates.bib,refsnelson.bib,RefCeline.bib}

\begin{thebibliography}{}
\makeatletter
\relax
\def\mn@urlcharsother{\let\do\@makeother \do\$\do\&\do\#\do\^\do\_\do\%\do\~}
\def\mn@doi{\begingroup\mn@urlcharsother \@ifnextchar [ {\mn@doi@}
  {\mn@doi@[]}}
\def\mn@doi@[#1]#2{\def\@tempa{#1}\ifx\@tempa\@empty \href
  {http://dx.doi.org/#2} {doi:#2}\else \href {http://dx.doi.org/#2} {#1}\fi
  \endgroup}
\def\mn@eprint#1#2{\mn@eprint@#1:#2::\@nil}
\def\mn@eprint@arXiv#1{\href {http://arxiv.org/abs/#1} {{\tt arXiv:#1}}}
\def\mn@eprint@dblp#1{\href {http://dblp.uni-trier.de/rec/bibtex/#1.xml}
  {dblp:#1}}
\def\mn@eprint@#1:#2:#3:#4\@nil{\def\@tempa {#1}\def\@tempb {#2}\def\@tempc
  {#3}\ifx \@tempc \@empty \let \@tempc \@tempb \let \@tempb \@tempa \fi \ifx
  \@tempb \@empty \def\@tempb {arXiv}\fi \@ifundefined
  {mn@eprint@\@tempb}{\@tempb:\@tempc}{\expandafter \expandafter \csname
  mn@eprint@\@tempb\endcsname \expandafter{\@tempc}}}

\bibitem[\protect\citeauthoryear{{Adelberger}, {Steidel}, {Shapley}  \&
  {Pettini}}{{Adelberger} et~al.}{2003}]{Adelberger+03}
{Adelberger} K.~L.,  {Steidel} C.~C.,  {Shapley} A.~E.,   {Pettini} M.,  2003,
  \mn@doi [\apj] {10.1086/345660}, \href
  {https://ui.adsabs.harvard.edu/abs/2003ApJ...584...45A} {584, 45}

\bibitem[\protect\citeauthoryear{{Aloisi}, {Savaglio}, {Heckman}, {Hoopes},
  {Leitherer}  \& {Sembach}}{{Aloisi} et~al.}{2003}]{aloisi2003}
{Aloisi} A.,  {Savaglio} S.,  {Heckman} T.~M.,  {Hoopes} C.~G.,  {Leitherer}
  C.,   {Sembach} K.~R.,  2003, \mn@doi [\apj] {10.1086/377496}, \href
  {https://ui.adsabs.harvard.edu/\#abs/2003ApJ...595..760A} {595, 760}

\bibitem[\protect\citeauthoryear{{Anderson} \& {Sunyaev}}{{Anderson} \&
  {Sunyaev}}{2016}]{anderson2016}
{Anderson} M.~E.,  {Sunyaev} R.,  2016, \mn@doi [\mnras]
  {10.1093/mnras/stw822}, \href
  {https://ui.adsabs.harvard.edu/\#abs/2016MNRAS.459.2806A} {459, 2806}

\bibitem[\protect\citeauthoryear{{Anderson} \& {Sunyaev}}{{Anderson} \&
  {Sunyaev}}{2018}]{anderson2018}
{Anderson} M.~E.,  {Sunyaev} R.,  2018, \mn@doi [\aap]
  {10.1051/0004-6361/201732510}, \href
  {https://ui.adsabs.harvard.edu/\#abs/2018A&A...617A.123A} {617, A123}

\bibitem[\protect\citeauthoryear{{Anderson}, {Bregman}, {Butler}  \&
  {Mullis}}{{Anderson} et~al.}{2009}]{Anderson+09}
{Anderson} M.~E.,  {Bregman} J.~N.,  {Butler} S.~C.,   {Mullis} C.~R.,  2009,
  \mn@doi [\apj] {10.1088/0004-637X/698/1/317}, \href
  {https://ui.adsabs.harvard.edu/abs/2009ApJ...698..317A} {698, 317}

\bibitem[\protect\citeauthoryear{{Andrews} \& {Martini}}{{Andrews} \&
  {Martini}}{2013}]{Andrews&Martini13}
{Andrews} B.~H.,  {Martini} P.,  2013, \mn@doi [\apj]
  {10.1088/0004-637X/765/2/140}, \href
  {http://adsabs.harvard.edu/abs/2013ApJ...765..140A} {765, 140}

\bibitem[\protect\citeauthoryear{{Angulo} \& {Hilbert}}{{Angulo} \&
  {Hilbert}}{2015}]{Angulo&Hilbert15}
{Angulo} R.~E.,  {Hilbert} S.,  2015, \mn@doi [MNRAS] {10.1093/mnras/stv050},
  \href {http://adsabs.harvard.edu/abs/2015MNRAS.448..364A} {448, 364}

\bibitem[\protect\citeauthoryear{{Angulo} \& {White}}{{Angulo} \&
  {White}}{2010}]{Angulo&White10}
{Angulo} R.~E.,  {White} S.~D.~M.,  2010, \mn@doi [MNRAS]
  {10.1111/j.1365-2966.2010.16459.x}, \href
  {http://adsabs.harvard.edu/abs/2010MNRAS.405..143A} {405, 143}

\bibitem[\protect\citeauthoryear{{Asada}, {Ohta}  \& {Maeda}}{{Asada}
  et~al.}{2021}]{Asada+21}
{Asada} Y.,  {Ohta} K.,   {Maeda} F.,  2021, \mn@doi [\apj]
  {10.3847/1538-4357/ac005a}, \href
  {https://ui.adsabs.harvard.edu/abs/2021ApJ...915...47A} {915, 47}

\bibitem[\protect\citeauthoryear{{Asplund}, {Grevesse}, {Sauval}  \&
  {Scott}}{{Asplund} et~al.}{2009}]{Asplund+09}
{Asplund} M.,  {Grevesse} N.,  {Sauval} A.~J.,   {Scott} P.,  2009, \mn@doi
  [\araa] {10.1146/annurev.astro.46.060407.145222}, \href
  {http://adsabs.harvard.edu/abs/2009ARA%26A..47..481A} {47, 481}

\bibitem[\protect\citeauthoryear{{Augustin} et~al.,}{{Augustin}
  et~al.}{2018}]{Augustin+18}
{Augustin} R.,  et~al., 2018, \mn@doi [\mnras] {10.1093/mnras/sty1287}, \href
  {https://ui.adsabs.harvard.edu/abs/2018MNRAS.478.3120A} {478, 3120}

\bibitem[\protect\citeauthoryear{{Ayromlou}, {Nelson}, {Yates}, {Kauffmann},
  {Renneby}  \& {White}}{{Ayromlou} et~al.}{2021}]{ayromlou20}
{Ayromlou} M.,  {Nelson} D.,  {Yates} R.~M.,  {Kauffmann} G.,  {Renneby} M.,
  {White} S. D.~M.,  2021, \mn@doi [\mnras] {10.1093/mnras/staa4011}, \href
  {https://ui.adsabs.harvard.edu/abs/2021MNRAS.502.1051A} {502, 1051}

\bibitem[\protect\citeauthoryear{{Baldi}, {Ettori}, {Molendi}, {Balestra},
  {Gastaldello}  \& {Tozzi}}{{Baldi} et~al.}{2012}]{Baldi+12}
{Baldi} A.,  {Ettori} S.,  {Molendi} S.,  {Balestra} I.,  {Gastaldello} F.,
  {Tozzi} P.,  2012, \mn@doi [\aap] {10.1051/0004-6361/201117836}, \href
  {https://ui.adsabs.harvard.edu/abs/2012A&A...537A.142B} {537, A142}

\bibitem[\protect\citeauthoryear{{Balestra}, {Tozzi}, {Ettori}, {Rosati},
  {Borgani}, {Mainieri}, {Norman}  \& {Viola}}{{Balestra}
  et~al.}{2007}]{Balestra+07}
{Balestra} I.,  {Tozzi} P.,  {Ettori} S.,  {Rosati} P.,  {Borgani} S.,
  {Mainieri} V.,  {Norman} C.,   {Viola} M.,  2007, \mn@doi [\aap]
  {10.1051/0004-6361:20065568}, \href
  {https://ui.adsabs.harvard.edu/abs/2007A&A...462..429B} {462, 429}

\bibitem[\protect\citeauthoryear{{Bellstedt} et~al.,}{{Bellstedt}
  et~al.}{2020}]{Bellstedt+20}
{Bellstedt} S.,  et~al., 2020, \mn@doi [\mnras] {10.1093/mnras/staa2620}, \href
  {https://ui.adsabs.harvard.edu/abs/2020MNRAS.498.5581B} {498, 5581}

\bibitem[\protect\citeauthoryear{{Berg} et~al.,}{{Berg} et~al.}{2019}]{berg19}
{Berg} M.~A.,  et~al., 2019, \mn@doi [\apj] {10.3847/1538-4357/ab378e}, \href
  {https://ui.adsabs.harvard.edu/abs/2019ApJ...883....5B} {883, 5}

\bibitem[\protect\citeauthoryear{{Berg} et~al.,}{{Berg} et~al.}{2021}]{Berg+21}
{Berg} T. A.~M.,  et~al., 2021, \mn@doi [\mnras] {10.1093/mnras/stab184}, \href
  {https://ui.adsabs.harvard.edu/abs/2021MNRAS.502.4009B} {502, 4009}

\bibitem[\protect\citeauthoryear{{Berry}, {Somerville}, {Haas}, {Gawiser},
  {Maller}, {Popping}  \& {Trager}}{{Berry} et~al.}{2014}]{Berry+14}
{Berry} M.,  {Somerville} R.~S.,  {Haas} M.~R.,  {Gawiser} E.,  {Maller} A.,
  {Popping} G.,   {Trager} S.~C.,  2014, \mn@doi [\mnras]
  {10.1093/mnras/stu613}, \href
  {https://ui.adsabs.harvard.edu/abs/2014MNRAS.441..939B} {441, 939}

\bibitem[\protect\citeauthoryear{{Berry}, {Somerville}, {Gawiser}, {Maller},
  {Popping}  \& {Trager}}{{Berry} et~al.}{2016}]{Berry+16}
{Berry} M.,  {Somerville} R.~S.,  {Gawiser} E.,  {Maller} A.~H.,  {Popping} G.,
    {Trager} S.~C.,  2016, \mn@doi [\mnras] {10.1093/mnras/stw231}, \href
  {https://ui.adsabs.harvard.edu/abs/2016MNRAS.458..531B} {458, 531}

\bibitem[\protect\citeauthoryear{{Bird}, {Vogelsberger}, {Haehnelt}, {Sijacki},
  {Genel}, {Torrey}, {Springel}  \& {Hernquist}}{{Bird} et~al.}{2014}]{bird14}
{Bird} S.,  {Vogelsberger} M.,  {Haehnelt} M.,  {Sijacki} D.,  {Genel} S.,
  {Torrey} P.,  {Springel} V.,   {Hernquist} L.,  2014, \mn@doi [\mnras]
  {10.1093/mnras/stu1923}, \href
  {http://adsabs.harvard.edu/abs/2014MNRAS.445.2313B} {445, 2313}

\bibitem[\protect\citeauthoryear{{Bischetti}, {Maiolino}, {Carniani}, {Fiore},
  {Piconcelli}  \& {Fluetsch}}{{Bischetti} et~al.}{2019}]{Bischetti+19}
{Bischetti} M.,  {Maiolino} R.,  {Carniani} S.,  {Fiore} F.,  {Piconcelli} E.,
   {Fluetsch} A.,  2019, \mn@doi [\aap] {10.1051/0004-6361/201833557}, \href
  {https://ui.adsabs.harvard.edu/abs/2019A&A...630A..59B} {630, A59}

\bibitem[\protect\citeauthoryear{{Blyth} et~al.,}{{Blyth}
  et~al.}{2016}]{Blyth+16}
{Blyth} S.,  et~al., 2016, in MeerKAT Science: On the Pathway to the SKA. p.~4

\bibitem[\protect\citeauthoryear{{Bouch{\'e}}, {Lehnert}  \&
  {P{\'e}roux}}{{Bouch{\'e}} et~al.}{2005}]{bouche2005}
{Bouch{\'e}} N.,  {Lehnert} M.~D.,   {P{\'e}roux} C.,  2005, \mn@doi [\mnras]
  {10.1111/j.1365-2966.2005.09570.x}, \href
  {https://ui.adsabs.harvard.edu/\#abs/2005MNRAS.364..319B} {364, 319}

\bibitem[\protect\citeauthoryear{{Bouch{\'e}}, {Lehnert}  \&
  {P{\'e}roux}}{{Bouch{\'e}} et~al.}{2006}]{bouche2006}
{Bouch{\'e}} N.,  {Lehnert} M.~D.,   {P{\'e}roux} C.,  2006, \mn@doi [\mnras]
  {10.1111/j.1745-3933.2005.00130.x}, \href
  {https://ui.adsabs.harvard.edu/\#abs/2006MNRAS.367L..16B} {367, L16}

\bibitem[\protect\citeauthoryear{{Bouch{\'e}}, {Lehnert}, {Aguirre},
  {P{\'e}roux}  \& {Bergeron}}{{Bouch{\'e}} et~al.}{2007}]{bouche2007}
{Bouch{\'e}} N.,  {Lehnert} M.~D.,  {Aguirre} A.,  {P{\'e}roux} C.,
  {Bergeron} J.,  2007, \mn@doi [\mnras] {10.1111/j.1365-2966.2007.11740.x},
  \href {https://ui.adsabs.harvard.edu/\#abs/2007MNRAS.378..525B} {378, 525}

\bibitem[\protect\citeauthoryear{Bowen, Jenkins, Pettini  \& Tripp}{Bowen
  et~al.}{2005}]{bowen2005}
Bowen D.,  Jenkins E.,  Pettini M.,   Tripp T.,  2005, ApJ, 635, 880

\bibitem[\protect\citeauthoryear{{Boylan-Kolchin}, {Springel}, {White},
  {Jenkins}  \& {Lemson}}{{Boylan-Kolchin} et~al.}{2009}]{Boylan-Kolchin+09}
{Boylan-Kolchin} M.,  {Springel} V.,  {White} S.~D.~M.,  {Jenkins} A.,
  {Lemson} G.,  2009, \mn@doi [MNRAS] {10.1111/j.1365-2966.2009.15191.x}, \href
  {http://adsabs.harvard.edu/abs/2009MNRAS.398.1150B} {398, 1150}

\bibitem[\protect\citeauthoryear{{Burchett} et~al.,}{{Burchett}
  et~al.}{2019}]{burchett2019}
{Burchett} J.~N.,  et~al., 2019, \mn@doi [\apjl] {10.3847/2041-8213/ab1f7f},
  \href {https://ui.adsabs.harvard.edu/abs/2019ApJ...877L..20B} {877, L20}

\bibitem[\protect\citeauthoryear{{Byrohl} et~al.,}{{Byrohl}
  et~al.}{2021}]{Byrohl+21}
{Byrohl} C.,  et~al., 2021, \mn@doi [\mnras] {10.1093/mnras/stab1958}, \href
  {https://ui.adsabs.harvard.edu/abs/2021MNRAS.506.5129B} {506, 5129}

\bibitem[\protect\citeauthoryear{{Christensen}, {M{\o}ller}, {Fynbo}  \&
  {Zafar}}{{Christensen} et~al.}{2014}]{Christensen+14}
{Christensen} L.,  {M{\o}ller} P.,  {Fynbo} J.~P.~U.,   {Zafar} T.,  2014,
  \mn@doi [\mnras] {10.1093/mnras/stu1726}, \href
  {https://ui.adsabs.harvard.edu/abs/2014MNRAS.445..225C} {445, 225}

\bibitem[\protect\citeauthoryear{{Cirasuolo} et~al.,}{{Cirasuolo}
  et~al.}{2020}]{Cirasuolo20}
{Cirasuolo} M.,  et~al., 2020, \mn@doi [The Messenger]
  {10.18727/0722-6691/5195}, \href
  {https://ui.adsabs.harvard.edu/abs/2020Msngr.180...10C} {180, 10}

\bibitem[\protect\citeauthoryear{{Crain} et~al.,}{{Crain}
  et~al.}{2015}]{crain15}
{Crain} R.~A.,  et~al., 2015, \mn@doi [\mnras] {10.1093/mnras/stv725}, \href
  {http://adsabs.harvard.edu/abs/2015MNRAS.450.1937C} {450, 1937}

\bibitem[\protect\citeauthoryear{{Curti} et~al.,}{{Curti}
  et~al.}{2020}]{Curti+20b}
{Curti} M.,  et~al., 2020, \mn@doi [\mnras] {10.1093/mnras/stz3379}, \href
  {https://ui.adsabs.harvard.edu/abs/2020MNRAS.492..821C} {492, 821}

\bibitem[\protect\citeauthoryear{{Dalla Vecchia} \& {Schaye}}{{Dalla Vecchia}
  \& {Schaye}}{2012}]{dallavecchia12}
{Dalla Vecchia} C.,  {Schaye} J.,  2012, \mn@doi [\mnras]
  {10.1111/j.1365-2966.2012.21704.x}, \href
  {https://ui.adsabs.harvard.edu/abs/2012MNRAS.426..140D} {426, 140}

\bibitem[\protect\citeauthoryear{{Dalton} et~al.,}{{Dalton}
  et~al.}{2012}]{Dalton12}
{Dalton} G.,  et~al., 2012, in {McLean} I.~S.,  {Ramsay} S.~K.,   {Takami} H.,
  eds,  Society of Photo-Optical Instrumentation Engineers (SPIE) Conference
  Series Vol. 8446, Ground-based and Airborne Instrumentation for Astronomy IV.
  p. 84460P, \mn@doi{10.1117/12.925950}

\bibitem[\protect\citeauthoryear{{De Cia}}{{De Cia}}{2018}]{decia2018a}
{De Cia} A.,  2018, \mn@doi [\aap] {10.1051/0004-6361/201833034}, \href
  {https://ui.adsabs.harvard.edu/#abs/2018A&A...613L...2D} {613, L2}

\bibitem[\protect\citeauthoryear{{De Cia}, {Ledoux}, {Mattsson}, {Petitjean},
  {Srianand}, {Gavignaud}  \& {Jenkins}}{{De Cia} et~al.}{2016}]{decia2016}
{De Cia} A.,  {Ledoux} C.,  {Mattsson} L.,  {Petitjean} P.,  {Srianand} R.,
  {Gavignaud} I.,   {Jenkins} E.~B.,  2016, \mn@doi [\aap]
  {10.1051/0004-6361/201527895}, \href
  {https://ui.adsabs.harvard.edu/#abs/2016A&A...596A..97D} {596, A97}

\bibitem[\protect\citeauthoryear{{De Cia}, {Ledoux}, {Petitjean}  \&
  {Savaglio}}{{De Cia} et~al.}{2018}]{De_Cia+18}
{De Cia} A.,  {Ledoux} C.,  {Petitjean} P.,   {Savaglio} S.,  2018, \mn@doi
  [\aap] {10.1051/0004-6361/201731970}, \href
  {https://ui.adsabs.harvard.edu/abs/2018A&A...611A..76D} {611, A76}

\bibitem[\protect\citeauthoryear{{De Rossi}, {Bower}, {Font}, {Schaye}  \&
  {Theuns}}{{De Rossi} et~al.}{2017}]{DeRossi+17}
{De Rossi} M.~E.,  {Bower} R.~G.,  {Font} A.~S.,  {Schaye} J.,   {Theuns} T.,
  2017, \mn@doi [\mnras] {10.1093/mnras/stx2158}, \href
  {https://ui.adsabs.harvard.edu/abs/2017MNRAS.472.3354D} {472, 3354}

\bibitem[\protect\citeauthoryear{{Di Gioia}, {Cristiani}, {De Lucia}  \&
  {Xie}}{{Di Gioia} et~al.}{2020}]{Di_Gioia+20}
{Di Gioia} S.,  {Cristiani} S.,  {De Lucia} G.,   {Xie} L.,  2020, \mn@doi
  [\mnras] {10.1093/mnras/staa2067}, \href
  {https://ui.adsabs.harvard.edu/abs/2020MNRAS.497.2469D} {497, 2469}

\bibitem[\protect\citeauthoryear{{Diemer} et~al.,}{{Diemer}
  et~al.}{2019}]{diemer19}
{Diemer} B.,  et~al., 2019, \mn@doi [\mnras] {10.1093/mnras/stz1323}, \href
  {https://ui.adsabs.harvard.edu/abs/2019MNRAS.487.1529D} {487, 1529}

\bibitem[\protect\citeauthoryear{{Doherty}, {Gil-Pons}, {Lau}, {Lattanzio}  \&
  {Siess}}{{Doherty} et~al.}{2014}]{doherty14}
{Doherty} C.~L.,  {Gil-Pons} P.,  {Lau} H. H.~B.,  {Lattanzio} J.~C.,   {Siess}
  L.,  2014, \mn@doi [\mnras] {10.1093/mnras/stt1877}, \href
  {https://ui.adsabs.harvard.edu/abs/2014MNRAS.437..195D} {437, 195}

\bibitem[\protect\citeauthoryear{{Donnari} et~al.,}{{Donnari}
  et~al.}{2019}]{donnari19}
{Donnari} M.,  et~al., 2019, \mn@doi [\mnras] {10.1093/mnras/stz712}, \href
  {https://ui.adsabs.harvard.edu/abs/2019MNRAS.485.4817D} {485, 4817}

\bibitem[\protect\citeauthoryear{{Dopita}, {Kewley}, {Sutherland}  \&
  {Nicholls}}{{Dopita} et~al.}{2016}]{Dopita+16}
{Dopita} M.~A.,  {Kewley} L.~J.,  {Sutherland} R.~S.,   {Nicholls} D.~C.,
  2016, \mn@doi [\apss] {10.1007/s10509-016-2657-8}, \href
  {http://adsabs.harvard.edu/abs/2016Ap%26SS.361...61D} {361, 61}

\bibitem[\protect\citeauthoryear{{Driver} et~al.,}{{Driver}
  et~al.}{2018}]{Driver+18}
{Driver} S.~P.,  et~al., 2018, \mn@doi [\mnras] {10.1093/mnras/stx2728}, \href
  {https://ui.adsabs.harvard.edu/abs/2018MNRAS.475.2891D} {475, 2891}

\bibitem[\protect\citeauthoryear{{Dvorkin}, {Silk}, {Vangioni}, {Petitjean}  \&
  {Olive}}{{Dvorkin} et~al.}{2015}]{dvorkin2015}
{Dvorkin} I.,  {Silk} J.,  {Vangioni} E.,  {Petitjean} P.,   {Olive} K.~A.,
  2015, \mn@doi [\mnras] {10.1093/mnrasl/slv085}, \href
  {http://adsabs.harvard.edu/abs/2015MNRAS.452L..36D} {452, L36}

\bibitem[\protect\citeauthoryear{{Erkal}, {Gnedin}  \& {Kravtsov}}{{Erkal}
  et~al.}{2012}]{Erkal+12}
{Erkal} D.,  {Gnedin} N.~Y.,   {Kravtsov} A.~V.,  2012, \mn@doi [\apj]
  {10.1088/0004-637X/761/1/54}, \href
  {https://ui.adsabs.harvard.edu/abs/2012ApJ...761...54E} {761, 54}

\bibitem[\protect\citeauthoryear{{Ettori}, {Baldi}, {Balestra}, {Gastaldello},
  {Molendi}  \& {Tozzi}}{{Ettori} et~al.}{2015}]{ettori2015}
{Ettori} S.,  {Baldi} A.,  {Balestra} I.,  {Gastaldello} F.,  {Molendi} S.,
  {Tozzi} P.,  2015, \mn@doi [\aap] {10.1051/0004-6361/201425470}, \href
  {https://ui.adsabs.harvard.edu/abs/2015A&A...578A..46E} {578, A46}

\bibitem[\protect\citeauthoryear{{Faucher-Gigu{\`e}re}}{{Faucher-Gigu{\`e}re}}{2020}]{Faucher-Giguere20}
{Faucher-Gigu{\`e}re} C.-A.,  2020, \mn@doi [\mnras] {10.1093/mnras/staa302},
  \href {https://ui.adsabs.harvard.edu/abs/2020MNRAS.493.1614F} {493, 1614}

\bibitem[\protect\citeauthoryear{{Faucher-Gigu{\`e}re}, {Lidz}, {Zaldarriaga}
  \& {Hernquist}}{{Faucher-Gigu{\`e}re} et~al.}{2009}]{fg09}
{Faucher-Gigu{\`e}re} C.-A.,  {Lidz} A.,  {Zaldarriaga} M.,   {Hernquist} L.,
  2009, \mn@doi [\apj] {10.1088/0004-637X/703/2/1416}, \href
  {http://adsabs.harvard.edu/abs/2009ApJ...703.1416F} {703, 1416}

\bibitem[\protect\citeauthoryear{{Ferrara}, {Scannapieco}  \&
  {Bergeron}}{{Ferrara} et~al.}{2005}]{ferrara2005}
{Ferrara} A.,  {Scannapieco} E.,   {Bergeron} J.,  2005, \mn@doi [\apj]
  {10.1086/498845}, \href
  {https://ui.adsabs.harvard.edu/\#abs/2005ApJ...634L..37F} {634, L37}

\bibitem[\protect\citeauthoryear{{Fishlock}, {Karakas}, {Lugaro}  \&
  {Yong}}{{Fishlock} et~al.}{2014}]{fishlock14}
{Fishlock} C.~K.,  {Karakas} A.~I.,  {Lugaro} M.,   {Yong} D.,  2014, \mn@doi
  [\apj] {10.1088/0004-637X/797/1/44}, \href
  {https://ui.adsabs.harvard.edu/abs/2014ApJ...797...44F} {797, 44}

\bibitem[\protect\citeauthoryear{{Frank}, {Pieri}, {Mathur}, {Danforth}  \&
  {Shull}}{{Frank} et~al.}{2018}]{frank2018}
{Frank} S.,  {Pieri} M.~M.,  {Mathur} S.,  {Danforth} C.~W.,   {Shull} J.~M.,
  2018, \mn@doi [\mnras] {10.1093/mnras/sty294}, \href
  {https://ui.adsabs.harvard.edu/abs/2018MNRAS.476.1356F} {476, 1356}

\bibitem[\protect\citeauthoryear{{Fresco}, {P{\'e}roux}, {Merloni},
  {Hamanowicz}  \& {Szakacs}}{{Fresco} et~al.}{2020}]{Fresco+20}
{Fresco} A.~Y.,  {P{\'e}roux} C.,  {Merloni} A.,  {Hamanowicz} A.,   {Szakacs}
  R.,  2020, \mn@doi [\mnras] {10.1093/mnras/staa2971}, \href
  {https://ui.adsabs.harvard.edu/abs/2020MNRAS.499.5230F} {499, 5230}

\bibitem[\protect\citeauthoryear{{Friis} et~al.,}{{Friis}
  et~al.}{2015}]{Friis+15}
{Friis} M.,  et~al., 2015, \mn@doi [\mnras] {10.1093/mnras/stv960}, \href
  {http://adsabs.harvard.edu/abs/2015MNRAS.451..167F} {451, 167}

\bibitem[\protect\citeauthoryear{{Fu}, {Guo}, {Kauffmann}  \& {Krumholz}}{{Fu}
  et~al.}{2010}]{Fu+10}
{Fu} J.,  {Guo} Q.,  {Kauffmann} G.,   {Krumholz} M.~R.,  2010, \mn@doi [MNRAS]
  {10.1111/j.1365-2966.2010.17342.x}, \href
  {http://adsabs.harvard.edu/abs/2010MNRAS.409..515F} {409, 515}

\bibitem[\protect\citeauthoryear{{Fu} et~al.,}{{Fu} et~al.}{2013}]{Fu+13}
{Fu} J.,  et~al., 2013, \mn@doi [\mnras] {10.1093/mnras/stt1117}, \href
  {https://ui.adsabs.harvard.edu/abs/2013MNRAS.434.1531F} {434, 1531}

\bibitem[\protect\citeauthoryear{{Furlong} et~al.,}{{Furlong}
  et~al.}{2015}]{Furlong+15}
{Furlong} M.,  et~al., 2015, \mn@doi [\mnras] {10.1093/mnras/stv852}, \href
  {https://ui.adsabs.harvard.edu/abs/2015MNRAS.450.4486F} {450, 4486}

\bibitem[\protect\citeauthoryear{{Fynbo} et~al.,}{{Fynbo}
  et~al.}{2013}]{Fynbo+13}
{Fynbo} J.~P.~U.,  et~al., 2013, \mn@doi [\mnras] {10.1093/mnras/stt1579},
  \href {https://ui.adsabs.harvard.edu/abs/2013MNRAS.436..361F} {436, 361}

\bibitem[\protect\citeauthoryear{{Gardner} et~al.,}{{Gardner}
  et~al.}{2006}]{Gardner06}
{Gardner} J.~P.,  et~al., 2006, \mn@doi [\ssr] {10.1007/s11214-006-8315-7},
  \href {https://ui.adsabs.harvard.edu/abs/2006SSRv..123..485G} {123, 485}

\bibitem[\protect\citeauthoryear{{Garratt-Smithson}, {Power}, {Lagos},
  {Stevens}, {Allison}  \& {Sadler}}{{Garratt-Smithson}
  et~al.}{2021}]{Garratt-Smithson+21}
{Garratt-Smithson} L.,  {Power} C.,  {Lagos} C. d.~P.,  {Stevens} A. R.~H.,
  {Allison} J.~R.,   {Sadler} E.~M.,  2021, \mn@doi [\mnras]
  {10.1093/mnras/staa3870}, \href
  {https://ui.adsabs.harvard.edu/abs/2021MNRAS.501.4396G} {501, 4396}

\bibitem[\protect\citeauthoryear{{Genel} et~al.,}{{Genel}
  et~al.}{2014}]{genel14}
{Genel} S.,  et~al., 2014, \mn@doi [\mnras] {10.1093/mnras/stu1654}, \href
  {http://adsabs.harvard.edu/abs/2014MNRAS.445..175G} {445, 175}

\bibitem[\protect\citeauthoryear{{Gibson}, {Pilkington}, {Brook}, {Stinson}  \&
  {Bailin}}{{Gibson} et~al.}{2013}]{Gibson+13}
{Gibson} B.~K.,  {Pilkington} K.,  {Brook} C.~B.,  {Stinson} G.~S.,   {Bailin}
  J.,  2013, \mn@doi [\aap] {10.1051/0004-6361/201321239}, \href
  {https://ui.adsabs.harvard.edu/abs/2013A&A...554A..47G} {554, A47}

\bibitem[\protect\citeauthoryear{{Gillman} et~al.,}{{Gillman}
  et~al.}{2021}]{Gillman+21}
{Gillman} S.,  et~al., 2021, \mn@doi [\mnras] {10.1093/mnras/staa3400}, \href
  {https://ui.adsabs.harvard.edu/abs/2021MNRAS.500.4229G} {500, 4229}

\bibitem[\protect\citeauthoryear{{Ginolfi} et~al.,}{{Ginolfi}
  et~al.}{2020}]{Ginolfi+20}
{Ginolfi} M.,  et~al., 2020, \mn@doi [\aap] {10.1051/0004-6361/201936872},
  \href {https://ui.adsabs.harvard.edu/abs/2020A&A...633A..90G} {633, A90}

\bibitem[\protect\citeauthoryear{{Gnedin} \& {Kravtsov}}{{Gnedin} \&
  {Kravtsov}}{2011}]{gnedin11}
{Gnedin} N.~Y.,  {Kravtsov} A.~V.,  2011, \mn@doi [\apj]
  {10.1088/0004-637X/728/2/88}, \href
  {https://ui.adsabs.harvard.edu/abs/2011ApJ...728...88G} {728, 88}

\bibitem[\protect\citeauthoryear{{Graham}, {Schady}  \& {Fruchter}}{{Graham}
  et~al.}{2019}]{Graham+19}
{Graham} J.~F.,  {Schady} P.,   {Fruchter} A.~S.,  2019, arXiv e-prints, \href
  {https://ui.adsabs.harvard.edu/abs/2019arXiv190402673G} {p. arXiv:1904.02673}

\bibitem[\protect\citeauthoryear{{Gruppioni} et~al.,}{{Gruppioni}
  et~al.}{2020}]{Gruppioni+20}
{Gruppioni} C.,  et~al., 2020, \mn@doi [\aap] {10.1051/0004-6361/202038487},
  \href {https://ui.adsabs.harvard.edu/abs/2020A&A...643A...8G} {643, A8}

\bibitem[\protect\citeauthoryear{{Guo}, {White}, {Angulo}, {Henriques},
  {Lemson}, {Boylan-Kolchin}, {Thomas}  \& {Short}}{{Guo}
  et~al.}{2013}]{Guo+13}
{Guo} Q.,  {White} S.,  {Angulo} R.~E.,  {Henriques} B.,  {Lemson} G.,
  {Boylan-Kolchin} M.,  {Thomas} P.,   {Short} C.,  2013, \mn@doi [\mnras]
  {10.1093/mnras/sts115}, \href
  {https://ui.adsabs.harvard.edu/abs/2013MNRAS.428.1351G} {428, 1351}

\bibitem[\protect\citeauthoryear{{Haardt} \& {Madau}}{{Haardt} \&
  {Madau}}{2001}]{haardt2001}
{Haardt} F.,  {Madau} P.,  2001, in Clusters of Galaxies and the High Redshift
  Universe Observed in X-rays. p.~64 (\mn@eprint {arXiv} {astro-ph/0106018})

\bibitem[\protect\citeauthoryear{{Hassan}, {Finlator}, {Dav{\'e}}, {Churchill}
  \& {Prochaska}}{{Hassan} et~al.}{2020}]{Hassan+20}
{Hassan} S.,  {Finlator} K.,  {Dav{\'e}} R.,  {Churchill} C.~W.,   {Prochaska}
  J.~X.,  2020, \mn@doi [\mnras] {10.1093/mnras/staa056}, \href
  {https://ui.adsabs.harvard.edu/abs/2020MNRAS.492.2835H} {492, 2835}

\bibitem[\protect\citeauthoryear{{Heintz}, {Watson}, {Oesch}, {Narayanan}  \&
  {Madden}}{{Heintz} et~al.}{2021}]{Heintz+21}
{Heintz} K.~E.,  {Watson} D.,  {Oesch} P.,  {Narayanan} D.,   {Madden} S.~C.,
  2021, arXiv e-prints, \href
  {https://ui.adsabs.harvard.edu/abs/2021arXiv210813442H} {p. arXiv:2108.13442}

\bibitem[\protect\citeauthoryear{{Hemler} et~al.,}{{Hemler}
  et~al.}{2020}]{Hemler+20}
{Hemler} Z.~S.,  et~al., 2020, arXiv e-prints, \href
  {https://ui.adsabs.harvard.edu/abs/2020arXiv200710993H} {p. arXiv:2007.10993}

\bibitem[\protect\citeauthoryear{{Henriques}, {Thomas}, {Oliver}  \&
  {Roseboom}}{{Henriques} et~al.}{2009}]{Henriques+09}
{Henriques} B. M.~B.,  {Thomas} P.~A.,  {Oliver} S.,   {Roseboom} I.,  2009,
  \mn@doi [\mnras] {10.1111/j.1365-2966.2009.14730.x}, \href
  {https://ui.adsabs.harvard.edu/abs/2009MNRAS.396..535H} {396, 535}

\bibitem[\protect\citeauthoryear{{Henriques}, {White}, {Thomas}, {Angulo},
  {Guo}, {Lemson}, {Springel}  \& {Overzier}}{{Henriques}
  et~al.}{2015}]{Henriques+15}
{Henriques} B.~M.~B.,  {White} S.~D.~M.,  {Thomas} P.~A.,  {Angulo} R.,  {Guo}
  Q.,  {Lemson} G.,  {Springel} V.,   {Overzier} R.,  2015, \mn@doi [MNRAS]
  {10.1093/mnras/stv705}, \href
  {http://adsabs.harvard.edu/abs/2015MNRAS.451.2663H} {451, 2663}

\bibitem[\protect\citeauthoryear{{Henriques}, {Yates}, {Fu}, {Guo},
  {Kauffmann}, {Srisawat}, {Thomas}  \& {White}}{{Henriques}
  et~al.}{2020}]{Henriques+20}
{Henriques} B. M.~B.,  {Yates} R.~M.,  {Fu} J.,  {Guo} Q.,  {Kauffmann} G.,
  {Srisawat} C.,  {Thomas} P.~A.,   {White} S. D.~M.,  2020, \mn@doi [\mnras]
  {10.1093/mnras/stz3233}, \href
  {https://ui.adsabs.harvard.edu/abs/2020MNRAS.491.5795H} {491, 5795}

\bibitem[\protect\citeauthoryear{{Hunt}, {Dayal}, {Magrini}  \&
  {Ferrara}}{{Hunt} et~al.}{2016}]{Hunt+16}
{Hunt} L.,  {Dayal} P.,  {Magrini} L.,   {Ferrara} A.,  2016, \mn@doi [\mnras]
  {10.1093/mnras/stw1993}, \href
  {http://adsabs.harvard.edu/abs/2016MNRAS.463.2002H} {463, 2002}

\bibitem[\protect\citeauthoryear{{Izotov}, {Guseva}, {Fricke}  \&
  {Henkel}}{{Izotov} et~al.}{2015}]{Izotov+15}
{Izotov} Y.~I.,  {Guseva} N.~G.,  {Fricke} K.~J.,   {Henkel} C.,  2015, \mn@doi
  [\mnras] {10.1093/mnras/stv1115}, \href
  {http://adsabs.harvard.edu/abs/2015MNRAS.451.2251I} {451, 2251}

\bibitem[\protect\citeauthoryear{{Jenkins}}{{Jenkins}}{2009}]{jenkins2009}
{Jenkins} E.~B.,  2009, \mn@doi [\apj] {10.1088/0004-637X/700/2/1299}, \href
  {http://adsabs.harvard.edu/abs/2009ApJ...700.1299J} {700, 1299}

\bibitem[\protect\citeauthoryear{{Jenkins} \& {Wallerstein}}{{Jenkins} \&
  {Wallerstein}}{2017}]{jenkins2017}
{Jenkins} E.~B.,  {Wallerstein} G.,  2017, \mn@doi [\apj]
  {10.3847/1538-4357/aa64d4}, \href
  {https://ui.adsabs.harvard.edu/#abs/2017ApJ...838...85J} {838, 85}

\bibitem[\protect\citeauthoryear{{Karakas}}{{Karakas}}{2010}]{karakas10}
{Karakas} A.~I.,  2010, \mn@doi [\mnras] {10.1111/j.1365-2966.2009.16198.x},
  \href {http://adsabs.harvard.edu/abs/2010MNRAS.403.1413K} {403, 1413}

\bibitem[\protect\citeauthoryear{{Katsianis} et~al.,}{{Katsianis}
  et~al.}{2017}]{Katsianis+17b}
{Katsianis} A.,  et~al., 2017, \mn@doi [\mnras] {10.1093/mnras/stx2020}, \href
  {https://ui.adsabs.harvard.edu/abs/2017MNRAS.472..919K} {472, 919}

\bibitem[\protect\citeauthoryear{{Katsianis} et~al.,}{{Katsianis}
  et~al.}{2020}]{Katsianis+20}
{Katsianis} A.,  et~al., 2020, \mn@doi [\mnras] {10.1093/mnras/staa157}, \href
  {https://ui.adsabs.harvard.edu/abs/2020MNRAS.492.5592K} {492, 5592}

\bibitem[\protect\citeauthoryear{{Kennicutt}}{{Kennicutt}}{1998}]{Kennicutt98a}
{Kennicutt} Robert~C. J.,  1998, \mn@doi [\apj] {10.1086/305588}, \href
  {https://ui.adsabs.harvard.edu/abs/1998ApJ...498..541K} {498, 541}

\bibitem[\protect\citeauthoryear{{Kewley} \& {Ellison}}{{Kewley} \&
  {Ellison}}{2008}]{Kewley&Ellison08}
{Kewley} L.~J.,  {Ellison} S.~L.,  2008, \mn@doi [\apj] {10.1086/587500}, \href
  {http://adsabs.harvard.edu/abs/2008ApJ...681.1183K} {681, 1183}

\bibitem[\protect\citeauthoryear{{Kistler}, {Y{\"u}ksel}, {Beacom}, {Hopkins}
  \& {Wyithe}}{{Kistler} et~al.}{2009}]{Kistler+09}
{Kistler} M.~D.,  {Y{\"u}ksel} H.,  {Beacom} J.~F.,  {Hopkins} A.~M.,
  {Wyithe} J. S.~B.,  2009, \mn@doi [\apjl] {10.1088/0004-637X/705/2/L104},
  \href {https://ui.adsabs.harvard.edu/abs/2009ApJ...705L.104K} {705, L104}

\bibitem[\protect\citeauthoryear{{Kobayashi}, {Umeda}, {Nomoto}, {Tominaga}  \&
  {Ohkubo}}{{Kobayashi} et~al.}{2006}]{kobayashi06}
{Kobayashi} C.,  {Umeda} H.,  {Nomoto} K.,  {Tominaga} N.,   {Ohkubo} T.,
  2006, \mn@doi [\apj] {10.1086/508914}, \href
  {https://ui.adsabs.harvard.edu/abs/2006ApJ...653.1145K} {653, 1145}

\bibitem[\protect\citeauthoryear{{Kobulnicky} \& {Kewley}}{{Kobulnicky} \&
  {Kewley}}{2004}]{Kobulnicky&Kewley04}
{Kobulnicky} H.~A.,  {Kewley} L.~J.,  2004, \mn@doi [\apj] {10.1086/425299},
  \href {http://adsabs.harvard.edu/abs/2004ApJ...617..240K} {617, 240}

\bibitem[\protect\citeauthoryear{{Koushan} et~al.,}{{Koushan}
  et~al.}{2021}]{Koushan+21}
{Koushan} S.,  et~al., 2021, \mn@doi [\mnras] {10.1093/mnras/stab540}, \href
  {https://ui.adsabs.harvard.edu/abs/2021MNRAS.503.2033K} {503, 2033}

\bibitem[\protect\citeauthoryear{Krogager et~al.,}{Krogager
  et~al.}{2013}]{krogager2013}
Krogager J.,  et~al., 2013, MNRAS, 433, 3091

\bibitem[\protect\citeauthoryear{{Krogager}, {M{\o}ller}, {Fynbo}  \&
  {Noterdaeme}}{{Krogager} et~al.}{2017}]{krogager2017}
{Krogager} J.-K.,  {M{\o}ller} P.,  {Fynbo} J.~P.~U.,   {Noterdaeme} P.,  2017,
  \mn@doi [MNRAS] {10.1093/mnras/stx1011}, \href
  {http://adsabs.harvard.edu/abs/2017MNRAS.469.2959K} {469, 2959}

\bibitem[\protect\citeauthoryear{{Krogager}, {Fynbo}, {M{\o}ller},
  {Noterdaeme}, {Heintz}  \& {Pettini}}{{Krogager} et~al.}{2019}]{Krogager+19}
{Krogager} J.-K.,  {Fynbo} J. P.~U.,  {M{\o}ller} P.,  {Noterdaeme} P.,
  {Heintz} K.~E.,   {Pettini} M.,  2019, \mn@doi [\mnras]
  {10.1093/mnras/stz1120}, \href
  {https://ui.adsabs.harvard.edu/abs/2019MNRAS.486.4377K} {486, 4377}

\bibitem[\protect\citeauthoryear{{Krogager}, {M{\o}ller}, {Christensen},
  {Noterdaeme}, {Fynbo}  \& {Freudling}}{{Krogager}
  et~al.}{2020}]{Krogager+20a}
{Krogager} J.-K.,  {M{\o}ller} P.,  {Christensen} L.~B.,  {Noterdaeme} P.,
  {Fynbo} J. P.~U.,   {Freudling} W.,  2020, \mn@doi [\mnras]
  {10.1093/mnras/staa1414}, \href
  {https://ui.adsabs.harvard.edu/abs/2020MNRAS.495.3014K} {495, 3014}

\bibitem[\protect\citeauthoryear{{Lagos}, {Baugh}, {Lacey}, {Benson}, {Kim}  \&
  {Power}}{{Lagos} et~al.}{2011}]{Lagos+11}
{Lagos} C. D.~P.,  {Baugh} C.~M.,  {Lacey} C.~G.,  {Benson} A.~J.,  {Kim}
  H.-S.,   {Power} C.,  2011, \mn@doi [\mnras]
  {10.1111/j.1365-2966.2011.19583.x}, \href
  {https://ui.adsabs.harvard.edu/abs/2011MNRAS.418.1649L} {418, 1649}

\bibitem[\protect\citeauthoryear{{Lagos}, {Baugh}, {Zwaan}, {Lacey},
  {Gonzalez-Perez}, {Power}, {Swinbank}  \& {van Kampen}}{{Lagos}
  et~al.}{2014}]{lagos2014}
{Lagos} C.~D.~P.,  {Baugh} C.~M.,  {Zwaan} M.~A.,  {Lacey} C.~G.,
  {Gonzalez-Perez} V.,  {Power} C.,  {Swinbank} A.~M.,   {van Kampen} E.,
  2014, \mn@doi [\mnras] {10.1093/mnras/stu266}, \href
  {https://ui.adsabs.harvard.edu/\#abs/2014MNRAS.440..920L} {440, 920}

\bibitem[\protect\citeauthoryear{Lebouteiller, Heap, Hubeny  \&
  Kunth}{Lebouteiller et~al.}{2013}]{lebouteiller2013}
Lebouteiller V.,  Heap S.,  Hubeny I.,   Kunth D.,  2013, A\&A, 553, 16

\bibitem[\protect\citeauthoryear{{Lecavelier des Etangs}, {D{\'e}sert},
  {Kunth}, {Vidal-Madjar}, {Callejo}, {Ferlet}, {H{\'e}brard}  \&
  {Lebouteiller}}{{Lecavelier des Etangs}
  et~al.}{2004}]{Lecavelier_des_Etangs+04}
{Lecavelier des Etangs} A.,  {D{\'e}sert} J.~M.,  {Kunth} D.,  {Vidal-Madjar}
  A.,  {Callejo} G.,  {Ferlet} R.,  {H{\'e}brard} G.,   {Lebouteiller} V.,
  2004, \mn@doi [\aap] {10.1051/0004-6361:20031518}, \href
  {https://ui.adsabs.harvard.edu/abs/2004A&A...413..131L} {413, 131}

\bibitem[\protect\citeauthoryear{{Leethochawalit}, {Jones}, {Ellis}, {Stark},
  {Richard}, {Zitrin}  \& {Auger}}{{Leethochawalit}
  et~al.}{2016}]{Leethochawalit+16}
{Leethochawalit} N.,  {Jones} T.~A.,  {Ellis} R.~S.,  {Stark} D.~P.,  {Richard}
  J.,  {Zitrin} A.,   {Auger} M.,  2016, \mn@doi [\apj]
  {10.3847/0004-637X/820/2/84}, \href
  {https://ui.adsabs.harvard.edu/abs/2016ApJ...820...84L} {820, 84}

\bibitem[\protect\citeauthoryear{{Leja} et~al.,}{{Leja} et~al.}{2019}]{leja19}
{Leja} J.,  et~al., 2019, \mn@doi [\apj] {10.3847/1538-4357/ab1d5a}, \href
  {https://ui.adsabs.harvard.edu/abs/2019ApJ...877..140L} {877, 140}

\bibitem[\protect\citeauthoryear{{Leja}, {Speagle}, {Johnson}, {Conroy}, {van
  Dokkum}  \& {Franx}}{{Leja} et~al.}{2020}]{leja20}
{Leja} J.,  {Speagle} J.~S.,  {Johnson} B.~D.,  {Conroy} C.,  {van Dokkum} P.,
   {Franx} M.,  2020, \mn@doi [\apj] {10.3847/1538-4357/ab7e27}, \href
  {https://ui.adsabs.harvard.edu/abs/2020ApJ...893..111L} {893, 111}

\bibitem[\protect\citeauthoryear{{Lemson} \& {Virgo Consortium}}{{Lemson} \&
  {Virgo Consortium}}{2006}]{Lemson+06}
{Lemson} G.,  {Virgo Consortium} t.,  2006, arXiv e-prints, \href
  {https://ui.adsabs.harvard.edu/abs/2006astro.ph..8019L} {pp
  astro--ph/0608019}

\bibitem[\protect\citeauthoryear{{Ma}, {Hopkins}, {Feldmann}, {Torrey},
  {Faucher-Gigu{\`e}re}  \& {Kere{\v{s}}}}{{Ma} et~al.}{2017}]{Ma+17}
{Ma} X.,  {Hopkins} P.~F.,  {Feldmann} R.,  {Torrey} P.,  {Faucher-Gigu{\`e}re}
  C.-A.,   {Kere{\v{s}}} D.,  2017, \mn@doi [\mnras] {10.1093/mnras/stx034},
  \href {https://ui.adsabs.harvard.edu/abs/2017MNRAS.466.4780M} {466, 4780}

\bibitem[\protect\citeauthoryear{{Madau} \& {Dickinson}}{{Madau} \&
  {Dickinson}}{2014}]{Madau&Dickinson14}
{Madau} P.,  {Dickinson} M.,  2014, \mn@doi [\araa]
  {10.1146/annurev-astro-081811-125615}, \href
  {https://ui.adsabs.harvard.edu/abs/2014ARA&A..52..415M} {52, 415}

\bibitem[\protect\citeauthoryear{{Maiolino} et~al.,}{{Maiolino}
  et~al.}{2008}]{Maiolino+08}
{Maiolino} R.,  et~al., 2008, \mn@doi [\aap] {10.1051/0004-6361:200809678},
  \href {http://adsabs.harvard.edu/abs/2008A%26A...488..463M} {488, 463}

\bibitem[\protect\citeauthoryear{{Marigo}}{{Marigo}}{2001}]{Marigo01}
{Marigo} P.,  2001, \mn@doi [\aap] {10.1051/0004-6361:20000247}, \href
  {https://ui.adsabs.harvard.edu/abs/2001A&A...370..194M} {370, 194}

\bibitem[\protect\citeauthoryear{{Marinacci} et~al.,}{{Marinacci}
  et~al.}{2018}]{marinacci18}
{Marinacci} F.,  et~al., 2018, \mn@doi [\mnras] {10.1093/mnras/sty2206}, \href
  {http://adsabs.harvard.edu/abs/2018MNRAS.480.5113M} {480, 5113}

\bibitem[\protect\citeauthoryear{{Marra} et~al.,}{{Marra}
  et~al.}{2021}]{Marra+21}
{Marra} R.,  et~al., 2021, arXiv e-prints, \href
  {https://ui.adsabs.harvard.edu/abs/2021arXiv210803254M} {p. arXiv:2108.03254}

\bibitem[\protect\citeauthoryear{{Martindale}, {Thomas}, {Henriques}  \&
  {Loveday}}{{Martindale} et~al.}{2017}]{Martindale+17}
{Martindale} H.,  {Thomas} P.~A.,  {Henriques} B.~M.,   {Loveday} J.,  2017,
  \mn@doi [\mnras] {10.1093/mnras/stx2131}, \href
  {https://ui.adsabs.harvard.edu/abs/2017MNRAS.472.1981M} {472, 1981}

\bibitem[\protect\citeauthoryear{{Matthews}, {Condon}, {Cotton}  \&
  {Mauch}}{{Matthews} et~al.}{2021}]{Matthews+21}
{Matthews} A.~M.,  {Condon} J.~J.,  {Cotton} W.~D.,   {Mauch} T.,  2021,
  \mn@doi [\apj] {10.3847/1538-4357/abfaf6}, \href
  {https://ui.adsabs.harvard.edu/abs/2021ApJ...914..126M} {914, 126}

\bibitem[\protect\citeauthoryear{{McAlpine} et~al.,}{{McAlpine}
  et~al.}{2016}]{McAlpine+16}
{McAlpine} S.,  et~al., 2016, \mn@doi [Astronomy and Computing]
  {10.1016/j.ascom.2016.02.004}, \href
  {https://ui.adsabs.harvard.edu/abs/2016A&C....15...72M} {15, 72}

\bibitem[\protect\citeauthoryear{{McDonald} et~al.,}{{McDonald}
  et~al.}{2016}]{McDonald+16}
{McDonald} M.,  et~al., 2016, \mn@doi [\apj] {10.3847/0004-637X/826/2/124},
  \href {https://ui.adsabs.harvard.edu/abs/2016ApJ...826..124M} {826, 124}

\bibitem[\protect\citeauthoryear{{McKee} \& {Krumholz}}{{McKee} \&
  {Krumholz}}{2010}]{McKee&Krumholz10}
{McKee} C.~F.,  {Krumholz} M.~R.,  2010, \mn@doi [\apj]
  {10.1088/0004-637X/709/1/308}, \href
  {https://ui.adsabs.harvard.edu/abs/2010ApJ...709..308M} {709, 308}

\bibitem[\protect\citeauthoryear{{Mitchell}, {Schaye}, {Bower}  \&
  {Crain}}{{Mitchell} et~al.}{2020}]{Mitchell+20a}
{Mitchell} P.~D.,  {Schaye} J.,  {Bower} R.~G.,   {Crain} R.~A.,  2020, \mn@doi
  [\mnras] {10.1093/mnras/staa938}, \href
  {https://ui.adsabs.harvard.edu/abs/2020MNRAS.494.3971M} {494, 3971}

\bibitem[\protect\citeauthoryear{M{\o}ller, Fynbo, Ledoux  \&
  Nilsson}{M{\o}ller et~al.}{2013}]{Moller+13}
M{\o}ller P.,  Fynbo J.,  Ledoux C.,   Nilsson K.,  2013, MNRAS, 430, 2680

\bibitem[\protect\citeauthoryear{{Mushotzky} \& {Loewenstein}}{{Mushotzky} \&
  {Loewenstein}}{1997}]{Mushotzky&Loewenstein97}
{Mushotzky} R.~F.,  {Loewenstein} M.,  1997, \mn@doi [\apjl] {10.1086/310651},
  \href {https://ui.adsabs.harvard.edu/abs/1997ApJ...481L..63M} {481, L63}

\bibitem[\protect\citeauthoryear{{Naiman} et~al.,}{{Naiman}
  et~al.}{2018}]{naiman18}
{Naiman} J.~P.,  et~al., 2018, \mn@doi [\mnras] {10.1093/mnras/sty618}, \href
  {http://adsabs.harvard.edu/abs/2018MNRAS.477.1206N} {477, 1206}

\bibitem[\protect\citeauthoryear{{Nandra} et~al.,}{{Nandra}
  et~al.}{2013}]{Kirpal13}
{Nandra} K.,  et~al., 2013, arXiv e-prints, \href
  {https://ui.adsabs.harvard.edu/abs/2013arXiv1306.2307N} {p. arXiv:1306.2307}

\bibitem[\protect\citeauthoryear{{Nelson} et~al.,}{{Nelson}
  et~al.}{2018a}]{nelson18a}
{Nelson} D.,  et~al., 2018a, \mn@doi [\mnras] {10.1093/mnras/stx3040}, \href
  {http://adsabs.harvard.edu/abs/2018MNRAS.475..624N} {475, 624}

\bibitem[\protect\citeauthoryear{{Nelson} et~al.,}{{Nelson}
  et~al.}{2018b}]{nelson18b}
{Nelson} D.,  et~al., 2018b, \mn@doi [\mnras] {10.1093/mnras/sty656}, \href
  {http://adsabs.harvard.edu/abs/2018MNRAS.477..450N} {477, 450}

\bibitem[\protect\citeauthoryear{{Nelson} et~al.,}{{Nelson}
  et~al.}{2019a}]{nelson19a}
{Nelson} D.,  et~al., 2019a, \mn@doi [Computational Astrophysics and Cosmology]
  {10.1186/s40668-019-0028-x}, \href
  {https://ui.adsabs.harvard.edu/abs/2019ComAC...6....2N} {6, 2}

\bibitem[\protect\citeauthoryear{{Nelson} et~al.,}{{Nelson}
  et~al.}{2019b}]{nelson19b}
{Nelson} D.,  et~al., 2019b, \mn@doi [\mnras] {10.1093/mnras/stz2306}, \href
  {https://ui.adsabs.harvard.edu/abs/2019MNRAS.490.3234N} {490, 3234}

\bibitem[\protect\citeauthoryear{{Nelson} et~al.,}{{Nelson}
  et~al.}{2021}]{Nelson+21}
{Nelson} E.~J.,  et~al., 2021, \mn@doi [\mnras] {10.1093/mnras/stab2131}, \href
  {https://ui.adsabs.harvard.edu/abs/2021MNRAS.508..219N} {508, 219}

\bibitem[\protect\citeauthoryear{{Nomoto}, {Hashimoto}, {Tsujimoto},
  {Thielemann}, {Kishimoto}, {Kubo}  \& {Nakasato}}{{Nomoto}
  et~al.}{1997}]{nomoto97}
{Nomoto} K.,  {Hashimoto} M.,  {Tsujimoto} T.,  {Thielemann} F.~K.,
  {Kishimoto} N.,  {Kubo} Y.,   {Nakasato} N.,  1997, \mn@doi [\nphysa]
  {10.1016/S0375-9474(97)00076-6}, \href
  {https://ui.adsabs.harvard.edu/abs/1997NuPhA.616...79N} {616, 79}

\bibitem[\protect\citeauthoryear{{Pagel}}{{Pagel}}{1999}]{pagel1999}
{Pagel} B.~E.~J.,  1999, arXiv e-prints, \href
  {https://ui.adsabs.harvard.edu/\#abs/1999astro.ph.11204P} {pp
  astro--ph/9911204}

\bibitem[\protect\citeauthoryear{{Patr{\'\i}cio} et~al.,}{{Patr{\'\i}cio}
  et~al.}{2019}]{Patricio+19}
{Patr{\'\i}cio} V.,  et~al., 2019, \mn@doi [\mnras] {10.1093/mnras/stz2114},
  \href {https://ui.adsabs.harvard.edu/abs/2019MNRAS.489..224P} {489, 224}

\bibitem[\protect\citeauthoryear{{Peeples} et~al.,}{{Peeples}
  et~al.}{2019}]{Peeples+19}
{Peeples} M.~S.,  et~al., 2019, \mn@doi [\apj] {10.3847/1538-4357/ab0654},
  \href {https://ui.adsabs.harvard.edu/abs/2019ApJ...873..129P} {873, 129}

\bibitem[\protect\citeauthoryear{{P{\'e}quignot}}{{P{\'e}quignot}}{2008}]{Pequignot08}
{P{\'e}quignot} D.,  2008, \mn@doi [\aap] {10.1051/0004-6361:20078344}, \href
  {https://ui.adsabs.harvard.edu/abs/2008A&A...478..371P} {478, 371}

\bibitem[\protect\citeauthoryear{{P{\'e}rez-R{\`a}fols}
  et~al.,}{{P{\'e}rez-R{\`a}fols} et~al.}{2018}]{Perez-Rafols+18a}
{P{\'e}rez-R{\`a}fols} I.,  et~al., 2018, \mn@doi [\mnras]
  {10.1093/mnras/stx2525}, \href
  {https://ui.adsabs.harvard.edu/abs/2018MNRAS.473.3019P} {473, 3019}

\bibitem[\protect\citeauthoryear{{P{\'e}roux} \& {Howk}}{{P{\'e}roux} \&
  {Howk}}{2020}]{Peroux&Howk20}
{P{\'e}roux} C.,  {Howk} J.~C.,  2020, \mn@doi [\araa]
  {10.1146/annurev-astro-021820-120014}, \href
  {https://ui.adsabs.harvard.edu/abs/2020ARA&A..58..363P} {58, 363}

\bibitem[\protect\citeauthoryear{{P{\'e}roux}, {Kulkarni}  \&
  {York}}{{P{\'e}roux} et~al.}{2014}]{peroux2014}
{P{\'e}roux} C.,  {Kulkarni} V.~P.,   {York} D.~G.,  2014, \mn@doi [\mnras]
  {10.1093/mnras/stt2084}, \href
  {https://ui.adsabs.harvard.edu/\#abs/2014MNRAS.437.3144P} {437, 3144}

\bibitem[\protect\citeauthoryear{{P{\'e}roux}, {Nelson}, {van de Voort},
  {Pillepich}, {Marinacci}, {Vogelsberger}  \& {Hernquist}}{{P{\'e}roux}
  et~al.}{2020}]{Peroux+20}
{P{\'e}roux} C.,  {Nelson} D.,  {van de Voort} F.,  {Pillepich} A.,
  {Marinacci} F.,  {Vogelsberger} M.,   {Hernquist} L.,  2020, \mn@doi [\mnras]
  {10.1093/mnras/staa2888}, \href
  {https://ui.adsabs.harvard.edu/abs/2020MNRAS.499.2462P} {499, 2462}

\bibitem[\protect\citeauthoryear{{Pettini}}{{Pettini}}{1999}]{pettini1999}
{Pettini} M.,  1999, in {Walsh} J.~R.,  {Rosa} M.~R.,  eds, Chemical Evolution
  from Zero to High Redshift. p.~233 (\mn@eprint {arXiv} {astro-ph/9902173})

\bibitem[\protect\citeauthoryear{{Pettini} \& {Dodorico}}{{Pettini} \&
  {Dodorico}}{1986}]{pettini1986}
{Pettini} M.,  {Dodorico} S.,  1986, \mn@doi [\apj] {10.1086/164721}, \href
  {https://ui.adsabs.harvard.edu/abs/1986ApJ...310..700P} {310, 700}

\bibitem[\protect\citeauthoryear{{Pettini} \& {Pagel}}{{Pettini} \&
  {Pagel}}{2004}]{Pettini&Pagel04}
{Pettini} M.,  {Pagel} B.~E.~J.,  2004, \mn@doi [\mnras]
  {10.1111/j.1365-2966.2004.07591.x}, \href
  {http://adsabs.harvard.edu/abs/2004MNRAS.348L..59P} {348, L59}

\bibitem[\protect\citeauthoryear{{Pillepich} et~al.,}{{Pillepich}
  et~al.}{2018a}]{pillepich18a}
{Pillepich} A.,  et~al., 2018a, \mn@doi [\mnras] {10.1093/mnras/stx2656}, \href
  {http://adsabs.harvard.edu/abs/2018MNRAS.473.4077P} {473, 4077}

\bibitem[\protect\citeauthoryear{{Pillepich} et~al.,}{{Pillepich}
  et~al.}{2018b}]{pillepich18b}
{Pillepich} A.,  et~al., 2018b, \mn@doi [\mnras] {10.1093/mnras/stx3112}, \href
  {http://adsabs.harvard.edu/abs/2018MNRAS.475..648P} {475, 648}

\bibitem[\protect\citeauthoryear{{Pillepich} et~al.,}{{Pillepich}
  et~al.}{2019}]{pillepich19}
{Pillepich} A.,  et~al., 2019, \mn@doi [\mnras] {10.1093/mnras/stz2338}, \href
  {https://ui.adsabs.harvard.edu/abs/2019MNRAS.490.3196P} {490, 3196}

\bibitem[\protect\citeauthoryear{{Planck Collaboration}}{{Planck
  Collaboration}}{2016}]{planck2015_xiii}
{Planck Collaboration} 2016, \mn@doi [\aap] {10.1051/0004-6361/201525830},
  \href {http://adsabs.harvard.edu/abs/2016A%26A...594A..13P} {594, A13}

\bibitem[\protect\citeauthoryear{{Planck Collaboration} et~al.,}{{Planck
  Collaboration} et~al.}{2014a}]{Planck14_I}
{Planck Collaboration} et~al., 2014a, \mn@doi [\aap]
  {10.1051/0004-6361/201321529}, \href
  {https://ui.adsabs.harvard.edu/abs/2014A&A...571A...1P} {571, A1}

\bibitem[\protect\citeauthoryear{{Planck Collaboration} et~al.,}{{Planck
  Collaboration} et~al.}{2014b}]{Planck14}
{Planck Collaboration} et~al., 2014b, \mn@doi [\aap]
  {10.1051/0004-6361/201321591}, \href
  {http://adsabs.harvard.edu/abs/2014A%26A...571A..16P} {571, A16}

\bibitem[\protect\citeauthoryear{Pontzen et~al.,}{Pontzen
  et~al.}{2008}]{pontzen2008}
Pontzen A.,  et~al., 2008, MNRAS, 390, 1349

\bibitem[\protect\citeauthoryear{{Popping} et~al.,}{{Popping}
  et~al.}{2019}]{popping19}
{Popping} G.,  et~al., 2019, \mn@doi [\apj] {10.3847/1538-4357/ab30f2}, \href
  {https://ui.adsabs.harvard.edu/abs/2019ApJ...882..137P} {882, 137}

\bibitem[\protect\citeauthoryear{{Portinari}, {Chiosi}  \&
  {Bressan}}{{Portinari} et~al.}{1998}]{Portinari+98}
{Portinari} L.,  {Chiosi} C.,   {Bressan} A.,  1998, \aap, \href
  {https://ui.adsabs.harvard.edu/abs/1998A&A...334..505P} {334, 505}

\bibitem[\protect\citeauthoryear{{Rafelski}, {Neeleman}, {Fumagalli}, {Wolfe}
  \& {Prochaska}}{{Rafelski} et~al.}{2014}]{rafelski2014}
{Rafelski} M.,  {Neeleman} M.,  {Fumagalli} M.,  {Wolfe} A.~M.,   {Prochaska}
  J.~X.,  2014, \mn@doi [\apj] {10.1088/2041-8205/782/2/L29}, \href
  {https://ui.adsabs.harvard.edu/#abs/2014ApJ...782L..29R} {782, L29}

\bibitem[\protect\citeauthoryear{{Rahmani} et~al.,}{{Rahmani}
  et~al.}{2016}]{rahmani2016}
{Rahmani} H.,  et~al., 2016, \mn@doi [\mnras] {10.1093/mnras/stw1965}, \href
  {http://adsabs.harvard.edu/abs/2016MNRAS.463..980R} {463, 980}

\bibitem[\protect\citeauthoryear{{Rahmati}, {Pawlik}, {Rai{\v c}evic}  \&
  {Schaye}}{{Rahmati} et~al.}{2013}]{rahmati13}
{Rahmati} A.,  {Pawlik} A.~H.,  {Rai{\v c}evic} M.,   {Schaye} J.,  2013,
  \mn@doi [\mnras] {10.1093/mnras/stt066}, \href
  {http://adsabs.harvard.edu/abs/2013MNRAS.430.2427R} {430, 2427}

\bibitem[\protect\citeauthoryear{{Rahmati}, {Schaye}, {Bower}, {Crain},
  {Furlong}, {Schaller}  \& {Theuns}}{{Rahmati} et~al.}{2015}]{Rahmati+15}
{Rahmati} A.,  {Schaye} J.,  {Bower} R.~G.,  {Crain} R.~A.,  {Furlong} M.,
  {Schaller} M.,   {Theuns} T.,  2015, \mn@doi [\mnras]
  {10.1093/mnras/stv1414}, \href
  {https://ui.adsabs.harvard.edu/abs/2015MNRAS.452.2034R} {452, 2034}

\bibitem[\protect\citeauthoryear{{Renzini}}{{Renzini}}{1998}]{renzini1998}
{Renzini} A.,  1998, arXiv e-prints, \href
  {https://ui.adsabs.harvard.edu/\#abs/1998astro.ph.10304R} {pp
  astro--ph/9810304}

\bibitem[\protect\citeauthoryear{{Rhodin}, {Krogager}, {Christensen},
  {Valentino}, {Heintz}, {M{\o}ller}, {Zafar}  \& {Fynbo}}{{Rhodin}
  et~al.}{2021}]{Rhodin+21}
{Rhodin} N.~H.~P.,  {Krogager} J.~K.,  {Christensen} L.,  {Valentino} F.,
  {Heintz} K.~E.,  {M{\o}ller} P.,  {Zafar} T.,   {Fynbo} J.~P.~U.,  2021,
  \mn@doi [\mnras] {10.1093/mnras/stab1691}, \href
  {https://ui.adsabs.harvard.edu/abs/2021MNRAS.506..546R} {506, 546}

\bibitem[\protect\citeauthoryear{{Richard} et~al.,}{{Richard}
  et~al.}{2019}]{Richard19}
{Richard} J.,  et~al., 2019, arXiv e-prints, \href
  {https://ui.adsabs.harvard.edu/abs/2019arXiv190601657R} {p. arXiv:1906.01657}

\bibitem[\protect\citeauthoryear{{Rowan-Robinson} et~al.,}{{Rowan-Robinson}
  et~al.}{2016}]{Rowan-Robinson+16}
{Rowan-Robinson} M.,  et~al., 2016, \mn@doi [\mnras] {10.1093/mnras/stw1169},
  \href {https://ui.adsabs.harvard.edu/abs/2016MNRAS.461.1100R} {461, 1100}

\bibitem[\protect\citeauthoryear{{Sanders} et~al.,}{{Sanders}
  et~al.}{2016}]{Sanders+16}
{Sanders} R.~L.,  et~al., 2016, \mn@doi [\apjl] {10.3847/2041-8205/825/2/L23},
  \href {https://ui.adsabs.harvard.edu/abs/2016ApJ...825L..23S} {825, L23}

\bibitem[\protect\citeauthoryear{{Sanders} et~al.,}{{Sanders}
  et~al.}{2021}]{Sanders+21}
{Sanders} R.~L.,  et~al., 2021, \mn@doi [\apj] {10.3847/1538-4357/abf4c1},
  \href {https://ui.adsabs.harvard.edu/abs/2021ApJ...914...19S} {914, 19}

\bibitem[\protect\citeauthoryear{{Schaye} et~al.,}{{Schaye}
  et~al.}{2010}]{schaye2010}
{Schaye} J.,  et~al., 2010, \mn@doi [MNRAS] {10.1111/j.1365-2966.2009.16029.x},
  \href {http://adsabs.harvard.edu/abs/2010MNRAS.402.1536S} {402, 1536}

\bibitem[\protect\citeauthoryear{{Schaye} et~al.,}{{Schaye}
  et~al.}{2015}]{schaye15}
{Schaye} J.,  et~al., 2015, \mn@doi [\mnras] {10.1093/mnras/stu2058}, \href
  {http://adsabs.harvard.edu/abs/2015MNRAS.446..521S} {446, 521}

\bibitem[\protect\citeauthoryear{Schulte-Ladbeck, Rao, Drozdovsky, Turnshek  \&
  Pettini}{Schulte-Ladbeck et~al.}{2004}]{schulte-ladbeck2004}
Schulte-Ladbeck R.,  Rao S.,  Drozdovsky I.,  Turnshek D.,   Pettini M.,  2004,
  ApJ, 600, 613

\bibitem[\protect\citeauthoryear{Schulte-Ladbeck, Konig, Miller, Hopkins,
  Drozdovsky, Turnshek  \& Hopp}{Schulte-Ladbeck
  et~al.}{2005}]{schulte-ladbeck2005}
Schulte-Ladbeck R.,  Konig B.,  Miller C.,  Hopkins A.,  Drozdovsky I.,
  Turnshek D.,   Hopp U.,  2005, ApJ, 625, L79

\bibitem[\protect\citeauthoryear{{Shull}, {Danforth}  \& {Tilton}}{{Shull}
  et~al.}{2014}]{shull2014}
{Shull} J.~M.,  {Danforth} C.~W.,   {Tilton} E.~M.,  2014, \mn@doi [\apj]
  {10.1088/0004-637X/796/1/49}, \href
  {https://ui.adsabs.harvard.edu/#abs/2014ApJ...796...49S} {796, 49}

\bibitem[\protect\citeauthoryear{{Spilker} et~al.,}{{Spilker}
  et~al.}{2020}]{Spilker+20}
{Spilker} J.~S.,  et~al., 2020, \mn@doi [\apj] {10.3847/1538-4357/abc4e6},
  \href {https://ui.adsabs.harvard.edu/abs/2020ApJ...905...86S} {905, 86}

\bibitem[\protect\citeauthoryear{{Spinelli}, {Zoldan}, {De Lucia}, {Xie}  \&
  {Viel}}{{Spinelli} et~al.}{2020}]{Spinelli+20}
{Spinelli} M.,  {Zoldan} A.,  {De Lucia} G.,  {Xie} L.,   {Viel} M.,  2020,
  \mn@doi [\mnras] {10.1093/mnras/staa604}, \href
  {https://ui.adsabs.harvard.edu/abs/2020MNRAS.493.5434S} {493, 5434}

\bibitem[\protect\citeauthoryear{{Springel}}{{Springel}}{2010}]{spr10}
{Springel} V.,  2010, \mn@doi [\mnras] {10.1111/j.1365-2966.2009.15715.x}, 401,
  791

\bibitem[\protect\citeauthoryear{{Springel} \& {Hernquist}}{{Springel} \&
  {Hernquist}}{2003}]{Springel&Hernquist03}
{Springel} V.,  {Hernquist} L.,  2003, \mn@doi [\mnras]
  {10.1046/j.1365-8711.2003.06206.x}, \href
  {https://ui.adsabs.harvard.edu/abs/2003MNRAS.339..289S} {339, 289}

\bibitem[\protect\citeauthoryear{{Springel}, {White}, {Tormen}  \&
  {Kauffmann}}{{Springel} et~al.}{2001}]{spr01}
{Springel} V.,  {White} S.~D.~M.,  {Tormen} G.,   {Kauffmann} G.,  2001,
  \mn@doi [\mnras] {10.1046/j.1365-8711.2001.04912.x}, \href
  {http://adsabs.harvard.edu/abs/2001MNRAS.328..726S} {328, 726}

\bibitem[\protect\citeauthoryear{{Springel}, {Di Matteo}  \&
  {Hernquist}}{{Springel} et~al.}{2005}]{spr05}
{Springel} V.,  {Di Matteo} T.,   {Hernquist} L.,  2005, \mn@doi [\mnras]
  {10.1111/j.1365-2966.2005.09238.x}, 361, 776

\bibitem[\protect\citeauthoryear{{Springel} et~al.,}{{Springel}
  et~al.}{2018}]{springel18}
{Springel} V.,  et~al., 2018, \mn@doi [\mnras] {10.1093/mnras/stx3304}, \href
  {http://adsabs.harvard.edu/abs/2018MNRAS.475..676S} {475, 676}

\bibitem[\protect\citeauthoryear{{Steidel}, {Erb}, {Shapley}, {Pettini},
  {Reddy}, {Bogosavljevi{\'c}}, {Rudie}  \& {Rakic}}{{Steidel}
  et~al.}{2010}]{steidel10}
{Steidel} C.~C.,  {Erb} D.~K.,  {Shapley} A.~E.,  {Pettini} M.,  {Reddy} N.,
  {Bogosavljevi{\'c}} M.,  {Rudie} G.~C.,   {Rakic} O.,  2010, \mn@doi [\apj]
  {10.1088/0004-637X/717/1/289}, \href
  {http://adsabs.harvard.edu/abs/2010ApJ...717..289S} {717, 289}

\bibitem[\protect\citeauthoryear{{Stevens} et~al.,}{{Stevens}
  et~al.}{2019}]{stevens19}
{Stevens} A. R.~H.,  et~al., 2019, \mn@doi [\mnras] {10.1093/mnras/sty3451},
  \href {https://ui.adsabs.harvard.edu/abs/2019MNRAS.483.5334S} {483, 5334}

\bibitem[\protect\citeauthoryear{{Stott} et~al.,}{{Stott}
  et~al.}{2014}]{Stott+14}
{Stott} J.~P.,  et~al., 2014, \mn@doi [\mnras] {10.1093/mnras/stu1343}, \href
  {https://ui.adsabs.harvard.edu/abs/2014MNRAS.443.2695S} {443, 2695}

\bibitem[\protect\citeauthoryear{{Sugahara}, {Ouchi}, {Harikane}, {Bouch{\'e}},
  {Mitchell}  \& {Blaizot}}{{Sugahara} et~al.}{2019}]{Sugahara+19}
{Sugahara} Y.,  {Ouchi} M.,  {Harikane} Y.,  {Bouch{\'e}} N.,  {Mitchell}
  P.~D.,   {Blaizot} J.,  2019, \mn@doi [\apj] {10.3847/1538-4357/ab49fe},
  \href {https://ui.adsabs.harvard.edu/abs/2019ApJ...886...29S} {886, 29}

\bibitem[\protect\citeauthoryear{{The Lynx Team}}{{The Lynx
  Team}}{2018}]{lynx2018}
{The Lynx Team} 2018, arXiv e-prints, \href
  {https://ui.adsabs.harvard.edu/abs/2018arXiv180909642T} {p. arXiv:1809.09642}

\bibitem[\protect\citeauthoryear{{The MSE Science Team} et~al.,}{{The MSE
  Science Team} et~al.}{2019}]{MSE19}
{The MSE Science Team} et~al., 2019, arXiv e-prints, \href
  {https://ui.adsabs.harvard.edu/abs/2019arXiv190404907T} {p. arXiv:1904.04907}

\bibitem[\protect\citeauthoryear{{Theuns}}{{Theuns}}{2021}]{Theuns21}
{Theuns} T.,  2021, \mn@doi [\mnras] {10.1093/mnras/staa3412}, \href
  {https://ui.adsabs.harvard.edu/abs/2021MNRAS.500.2741T} {500, 2741}

\bibitem[\protect\citeauthoryear{{Thielemann} et~al.,}{{Thielemann}
  et~al.}{2003}]{Thielemann+03}
{Thielemann} F.~K.,  et~al., 2003, in {Hillebrandt} W.,  {Leibundgut} B.,  eds,
  From Twilight to Highlight: The Physics of Supernovae. p.~331,
  \mn@doi{10.1007/10828549_46}

\bibitem[\protect\citeauthoryear{{Th{\"o}ne} et~al.,}{{Th{\"o}ne}
  et~al.}{2013}]{Thoene+13}
{Th{\"o}ne} C.~C.,  et~al., 2013, \mn@doi [\mnras] {10.1093/mnras/sts303},
  \href {https://ui.adsabs.harvard.edu/abs/2013MNRAS.428.3590T} {428, 3590}

\bibitem[\protect\citeauthoryear{{Torrey} et~al.,}{{Torrey}
  et~al.}{2019}]{Torrey+19}
{Torrey} P.,  et~al., 2019, \mn@doi [\mnras] {10.1093/mnras/stz243}, \href
  {https://ui.adsabs.harvard.edu/abs/2019MNRAS.484.5587T} {484, 5587}

\bibitem[\protect\citeauthoryear{{Tozzi}, {Rosati}, {Ettori}, {Borgani},
  {Mainieri}  \& {Norman}}{{Tozzi} et~al.}{2003}]{Tozzi+03}
{Tozzi} P.,  {Rosati} P.,  {Ettori} S.,  {Borgani} S.,  {Mainieri} V.,
  {Norman} C.,  2003, \mn@doi [\apj] {10.1086/376731}, \href
  {https://ui.adsabs.harvard.edu/abs/2003ApJ...593..705T} {593, 705}

\bibitem[\protect\citeauthoryear{{Tremonti} et~al.,}{{Tremonti}
  et~al.}{2004}]{Tremonti+04}
{Tremonti} C.~A.,  et~al., 2004, \mn@doi [\apj] {10.1086/423264}, \href
  {http://adsabs.harvard.edu/abs/2004ApJ...613..898T} {613, 898}

\bibitem[\protect\citeauthoryear{{Utomo}, {Kriek}, {Labb{\'e}}, {Conroy}  \&
  {Fumagalli}}{{Utomo} et~al.}{2014}]{Utomo+14}
{Utomo} D.,  {Kriek} M.,  {Labb{\'e}} I.,  {Conroy} C.,   {Fumagalli} M.,
  2014, \mn@doi [\apjl] {10.1088/2041-8205/783/2/L30}, \href
  {https://ui.adsabs.harvard.edu/abs/2014ApJ...783L..30U} {783, L30}

\bibitem[\protect\citeauthoryear{{Villaescusa-Navarro}
  et~al.,}{{Villaescusa-Navarro} et~al.}{2018}]{villaescusa-navarro2018}
{Villaescusa-Navarro} F.,  et~al., 2018, \mn@doi [\apj]
  {10.3847/1538-4357/aadba0}, \href
  {http://adsabs.harvard.edu/abs/2018ApJ...866..135V} {866, 135}

\bibitem[\protect\citeauthoryear{{Vladilo}, {Bonifacio}, {Centuri{\'o}n}  \&
  {Molaro}}{{Vladilo} et~al.}{2000}]{Vladilo+00}
{Vladilo} G.,  {Bonifacio} P.,  {Centuri{\'o}n} M.,   {Molaro} P.,  2000,
  \mn@doi [\apj] {10.1086/317110}, \href
  {https://ui.adsabs.harvard.edu/abs/2000ApJ...543...24V} {543, 24}

\bibitem[\protect\citeauthoryear{{Vogelsberger}, {Genel}, {Sijacki}, {Torrey},
  {Springel}  \& {Hernquist}}{{Vogelsberger} et~al.}{2013}]{vog13}
{Vogelsberger} M.,  {Genel} S.,  {Sijacki} D.,  {Torrey} P.,  {Springel} V.,
  {Hernquist} L.,  2013, \mn@doi [\mnras] {10.1093/mnras/stt1789}, \href
  {http://adsabs.harvard.edu/abs/2013MNRAS.436.3031V} {436, 3031}

\bibitem[\protect\citeauthoryear{{Wang}, {Pearson}, {Cowley}, {Trayford},
  {B{\'e}thermin}, {Gruppioni}, {Hurley}  \& {Micha{\l}owski}}{{Wang}
  et~al.}{2019}]{Wang+19}
{Wang} L.,  {Pearson} W.~J.,  {Cowley} W.,  {Trayford} J.~W.,  {B{\'e}thermin}
  M.,  {Gruppioni} C.,  {Hurley} P.,   {Micha{\l}owski} M.~J.,  2019, \mn@doi
  [\aap] {10.1051/0004-6361/201834093}, \href
  {https://ui.adsabs.harvard.edu/abs/2019A&A...624A..98W} {624, A98}

\bibitem[\protect\citeauthoryear{{Weinberger} et~al.,}{{Weinberger}
  et~al.}{2017}]{weinberger17}
{Weinberger} R.,  et~al., 2017, \mn@doi [\mnras] {10.1093/mnras/stw2944}, \href
  {http://adsabs.harvard.edu/abs/2017MNRAS.465.3291W} {465, 3291}

\bibitem[\protect\citeauthoryear{{Wiersma}, {Schaye}  \& {Smith}}{{Wiersma}
  et~al.}{2009}]{wiersma09}
{Wiersma} R.~P.~C.,  {Schaye} J.,   {Smith} B.~D.,  2009, \mn@doi [\mnras]
  {10.1111/j.1365-2966.2008.14191.x}, \href
  {http://adsabs.harvard.edu/abs/2009MNRAS.393...99W} {393, 99}

\bibitem[\protect\citeauthoryear{{Wilkins}, {Lovell}  \& {Stanway}}{{Wilkins}
  et~al.}{2019}]{Wilkins+19}
{Wilkins} S.~M.,  {Lovell} C.~C.,   {Stanway} E.~R.,  2019, \mn@doi [\mnras]
  {10.1093/mnras/stz2894}, \href
  {https://ui.adsabs.harvard.edu/abs/2019MNRAS.490.5359W} {490, 5359}

\bibitem[\protect\citeauthoryear{{Wolfe}, {Gawiser}  \& {Prochaska}}{{Wolfe}
  et~al.}{2005}]{wolfe2005}
{Wolfe} A.~M.,  {Gawiser} E.,   {Prochaska} J.~X.,  2005, \mn@doi [\araa]
  {10.1146/annurev.astro.42.053102.133950}, \href
  {http://adsabs.harvard.edu/abs/2005ARA%26A..43..861W} {43, 861}

\bibitem[\protect\citeauthoryear{{Xie}, {De Lucia}, {Hirschmann}, {Fontanot}
  \& {Zoldan}}{{Xie} et~al.}{2017}]{Xie+17}
{Xie} L.,  {De Lucia} G.,  {Hirschmann} M.,  {Fontanot} F.,   {Zoldan} A.,
  2017, \mn@doi [\mnras] {10.1093/mnras/stx889}, \href
  {https://ui.adsabs.harvard.edu/abs/2017MNRAS.469..968X} {469, 968}

\bibitem[\protect\citeauthoryear{{Yates}, {Henriques}, {Thomas}, {Kauffmann},
  {Johansson}  \& {White}}{{Yates} et~al.}{2013}]{Yates+13}
{Yates} R.~M.,  {Henriques} B.,  {Thomas} P.~A.,  {Kauffmann} G.,  {Johansson}
  J.,   {White} S. D.~M.,  2013, \mn@doi [\mnras] {10.1093/mnras/stt1542},
  \href {https://ui.adsabs.harvard.edu/abs/2013MNRAS.435.3500Y} {435, 3500}

\bibitem[\protect\citeauthoryear{{Yates}, {Schady}, {Chen}, {Schweyer}  \&
  {Wiseman}}{{Yates} et~al.}{2020}]{Yates+20}
{Yates} R.~M.,  {Schady} P.,  {Chen} T.~W.,  {Schweyer} T.,   {Wiseman} P.,
  2020, \mn@doi [\aap] {10.1051/0004-6361/201936506}, \href
  {https://ui.adsabs.harvard.edu/abs/2020A&A...634A.107Y} {634, A107}

\bibitem[\protect\citeauthoryear{{Yates}, {Henriques}, {Fu}, {Kauffmann},
  {Thomas}, {Guo}, {White}  \& {Schady}}{{Yates} et~al.}{2021}]{Yates+21a}
{Yates} R.~M.,  {Henriques} B. M.~B.,  {Fu} J.,  {Kauffmann} G.,  {Thomas}
  P.~A.,  {Guo} Q.,  {White} S. D.~M.,   {Schady} P.,  2021, \mn@doi [\mnras]
  {10.1093/mnras/stab741}, \href
  {https://ui.adsabs.harvard.edu/abs/2021MNRAS.503.4474Y} {503, 4474}

\bibitem[\protect\citeauthoryear{{Zafar}, {P{\'e}roux}, {Popping}, {Milliard},
  {Deharveng}  \& {Frank}}{{Zafar} et~al.}{2013}]{Zafar+13}
{Zafar} T.,  {P{\'e}roux} C.,  {Popping} A.,  {Milliard} B.,  {Deharveng}
  J.~M.,   {Frank} S.,  2013, \mn@doi [\aap] {10.1051/0004-6361/201321154},
  \href {https://ui.adsabs.harvard.edu/abs/2013A&A...556A.141Z} {556, A141}

\bibitem[\protect\citeauthoryear{{Zahid}, {Dima}, {Kudritzki}, {Kewley},
  {Geller}, {Hwang}, {Silverman}  \& {Kashino}}{{Zahid}
  et~al.}{2014}]{Zahid+14}
{Zahid} H.~J.,  {Dima} G.~I.,  {Kudritzki} R.-P.,  {Kewley} L.~J.,  {Geller}
  M.~J.,  {Hwang} H.~S.,  {Silverman} J.~D.,   {Kashino} D.,  2014, \mn@doi
  [\apj] {10.1088/0004-637X/791/2/130}, \href
  {http://adsabs.harvard.edu/abs/2014ApJ...791..130Z} {791, 130}

\bibitem[\protect\citeauthoryear{{Zavala} et~al.,}{{Zavala}
  et~al.}{2021}]{Zavala+21}
{Zavala} J.~A.,  et~al., 2021, \mn@doi [\apj] {10.3847/1538-4357/abdb27}, \href
  {https://ui.adsabs.harvard.edu/abs/2021ApJ...909..165Z} {909, 165}

\bibitem[\protect\citeauthoryear{{Zhao}, {Xu}, {Katsianis}  \& {Yang}}{{Zhao}
  et~al.}{2020}]{Zhao+20}
{Zhao} P.,  {Xu} H.,  {Katsianis} A.,   {Yang} X.-H.,  2020, \mn@doi [Research
  in Astronomy and Astrophysics] {10.1088/1674-4527/20/12/195}, \href
  {https://ui.adsabs.harvard.edu/abs/2020RAA....20..195Z} {20, 195}

\bibitem[\protect\citeauthoryear{{Zwaan}, {Meyer}, {Staveley-Smith}  \&
  {Webster}}{{Zwaan} et~al.}{2005}]{Zwaan+05}
{Zwaan} M.~A.,  {Meyer} M.~J.,  {Staveley-Smith} L.,   {Webster} R.~L.,  2005,
  \mn@doi [\mnras] {10.1111/j.1745-3933.2005.00029.x}, \href
  {https://ui.adsabs.harvard.edu/abs/2005MNRAS.359L..30Z} {359, L30}

\bibitem[\protect\citeauthoryear{{de Jong} et~al.,}{{de Jong}
  et~al.}{2019}]{deJong19}
{de Jong} R.~S.,  et~al., 2019, \mn@doi [The Messenger]
  {10.18727/0722-6691/5117}, \href
  {https://ui.adsabs.harvard.edu/abs/2019Msngr.175....3D} {175, 3}

\makeatother
\end{thebibliography}



\label{lastpage}
\end{document}